\documentclass[aip,jcp,amsmath,amssymb,floatfix,citeautoscript,reprint]{revtex4-1}

\usepackage{graphicx}
\usepackage{dcolumn}
\usepackage{bm}

\usepackage{braket}
\usepackage[usenames,dvipsnames]{color}

\usepackage{ragged2e}
\usepackage{times}
\usepackage{txfonts}

\begin{document}

\title{A derivation of the conditions under which bosonic operators exactly capture fermionic structure and dynamics}

\author{Andr\'es Montoya-Castillo}
\email{Andres.MontoyaCastillo@colorado.edu}
\affiliation{Department of Chemistry, University of Colorado, Boulder, Boulder, Colorado, 80309, USA}

\author{Thomas E. Markland}
\email{tmarkland@stanford.edu}
\affiliation{Department of Chemistry, Stanford University, Stanford, California, 94305, USA}

\date{\today}

\begin{abstract}
The dynamics of many-body fermionic systems are important in problems ranging from catalytic reactions at electrochemical surfaces, to transport through nanojunctions, and offer a prime target for quantum computing applications. Here we derive the set of conditions under which fermionic operators can be exactly replaced by bosonic operators that render the problem amenable to a large toolbox of dynamical methods while still capturing the correct dynamics of the $n$-body operators. Importantly, our analysis offers a simple guide on how one can exploit these simple maps to calculate nonequilibrium and equilibrium single- and multi-time correlation functions essential in describing transport and spectroscopy. We use this to rigorously analyze and delineate the applicability of simple yet effective Cartesian maps that have been shown to correctly capture the correct fermionic dynamics in select models of nanoscopic transport. We illustrate our analytical results with exact simulations of the resonant level model. Our work provides new insights as to when one can leverage the simplicity of bosonic maps to simulate the dynamics of many-electron systems, especially those where an atomistic representation of nuclear interactions becomes essential.
\end{abstract}

\maketitle

\section{Introduction}

Dynamical processes involving many electrons are ubiquitous. These range from magnetism and superconductivity \cite{Auerbach1998, dagottoCorrelatedElectronsHightemperature1994, orensteinAdvancesPhysicsHighTemperature2000, leeDopingMottInsulator2006}, to catalytic reactions at electrochemical interfaces \cite{chidseyFreeEnergyTemperature1991, sehCombiningTheoryExperiment2017, warburtonTheoreticalModelingElectrochemical2022, santosModelsElectronTransfer2022} and at molecular centers \cite{miglioreBiochemistryTheoryProtonCoupled2014, yulyElectronBifurcationProgress2019, pannwitzProtoncoupledMultielectronTransfer2019, rutledgeElectronTransferNitrogenase2020}, electric transport in bulk systems as well as through nanojunctions \cite{Evers2019, CohenGalperin2020}, and quantum computing \cite{atiaFastforwardingHamiltoniansExponentially2017, lammSimulationNonequilibriumDynamics2018,sunQuantumComputationFiniteTemperature2021, oftelieConstantdepthCircuitsDynamic2022}. However, because the Hilbert space of many-fermion problems scales exponentially with the number of available single-particle states, these systems pose unique challenges to existing theories and simulation methodologies. Trajectory-based path integral-based approaches and quantum-classical theories provide an approach to potentially alleviate this problem. These approaches offer a hierarchy of exact and approximate solutions to the dynamics that provide tradeoffs in accuracy and efficiency that have been successfully applied to problems ranging from excitation energy transport in molecular systems \cite{Kapral2015, Lee2016b}, to quantum optics and cavity electrodynamics \cite{QuantumOpticsInPhaseSpace, gardinerQuantumWorldUltraCold2014}, and interacting spin phenomena (e.g., frustration and magnetism) and in quantum information science \cite{Polkovnikov2010, Schachenmayer2015a, Swingle2018}. At the heart of these theories is the ability to cast the Hamiltonian and observables in terms of continuous degrees of freedom, such as action-angle and Cartesian variables. Once a problem can be articulated in terms of such variables, one can then apply the rich hierarchy of semi- and quantum-classical methods to these problems.

To create a robust simulation methodology for many-fermion problems based on the quantum-classical hierarchy, it is essential to have either an exact quantum mechanical map connecting fermionic operators and continuous variables from which one can devise approximate solution schemes, or an approximate map where the limits of applicability are clearly known. We have developed an exact quantum mechanical map for individual fermionic creation and annihilation operators that allows one to map any many-fermion problem to bosonic operators, which can then be cast in terms of Cartesian variables \cite{Montoya2018}. While this map is exact and is therefore guaranteed to maintain the exact structure and dynamics of the original fermionic problem, it also presents challenges from the quantum-classical perspective, including nonlocal operators that encode anticommutivity and the fact that each fermionic degree of freedom is associated with two correlated bosonic ones. It is thus desirable to have a controlled means to map many-fermion problems to a Cartesian representation that avoids the complexities of our previous map and is compatible with systematically improvable quantum-classical theories.

Recently a series of physically motivated quasiclassical Cartesian maps \cite{Li2012, Li2014b, Levy2019} have been suggested and shown numerically to perform well in reproducing the dynamics of model problems. These maps provide expressions either for individual fermionic creation and annihilation operators \cite{Li2014b} or their quadratic products \cite{Li2012, Levy2019} in terms of Cartesian variables. However, each approach is beset by specific difficulties. In the former case, as we show here, the previously proposed map results in operators that obey bosonic, rather than fermionic, commutation relations. In the latter, the lack of a well defined map for single creation and annihilation operators obscures the physical basis necessary for the application of a broad class of quantum-classical methods \cite{Sun1997, Muller1998, Muller1999, Wang1998b, Sun1998, Thoss1999, Thoss2000, Volobuev2000, Coronado2001, Liao2002, Shi2004a, Bonella2005, Ananth2007, Dunkel2008, Kim2008, Ananth2010, Huo2011, Hsieh2012, Kelly2012, Ananth2013, Richardson2013, Hele2016, Chowdhury2017, Church2018}. However, despite violating the fundamental anticommutivity of individual fermionic operators, these quasiclassical maps have been shown numerically to accurately capture some time-dependent observables for several commonly used models of nanoscopic transport \cite{Li2012, Li2013a, Li2014a, Li2014b, Levy2019}.
In addition, recent work \cite{Liu2017, sunBosonicPerspectiveClassical2021} has established that use of the Meyer-Miller-Stock-Thoss map \cite{Meyer1979, Stock1997}, traditionally used to replace outer products of discrete states with bosonic degrees of freedom, when combined with a classical evolution accurately captures the quantum dynamics of the one-body density of non-interacting fermionic Hamiltonians, provided the system starts from an occupied or unoccupied state. 

This motivates the fundamental question: when do maps that directly replace fermionic creation and annihilation operators $\{ \hat{c}_j^{\dagger}, \hat{c}_j\}$ with bosonic ones $\{ \hat{b}_j^{\dagger}, \hat{b}_j\}$, work? 

Here using exact quantum mechanical arguments we analytically establish the specific sets of conditions under which this seemingly naive map is guaranteed to provide the \emph{exact} matrix elements and dynamical observables for many-fermion problems. In particular, we demonstrate that the bosonic representation permits the exact calculation of the diagonal matrix elements of:
\begin{enumerate}
    \item Static operators where the resulting string of bosonic operators are at most an even permutation away from being pairwise-ordered, i.e., $\prod_{j}\hat{c}_j^{\dagger}\hat{c}_j$ or $\prod_{j}\hat{c}_j\hat{c}_j^{\dagger}$.
    
    \item Dynamical operators subject to a quadratic Hamiltonian consisting of, at most, the product of two creation-annihilation pairs in similar order, i.e., $\hat{c}_j^{\dagger}(t)\hat{c}_k(t)\hat{c}_l^{\dagger}(t)\hat{c}_m(t)$ or $\hat{c}_j(t)\hat{c}^{\dagger}_k(t)\hat{c}_l(t)\hat{c}^{\dagger}_m(t)$.
\end{enumerate}
In doing this, we also derive the types of observables, ranging from static expectation values to single and multi-time correlation functions, that one can correctly capture using such approximate maps. By considering the classical limit of our analytic results, we are able to show when the classical dynamics of mapped bosonic systems correctly capture the exact fermionic result. Finally, we demonstrate the validity of our analytical insights with numerical results for the transport characteristics of the resonant level model.

Our analysis provides insights as to the origin of the success of previous maps which, at first inspection, could have been expected to violate important properties of fermionic algebra; elucidates why the classical limit of these maps is capable of reproducing the exact quantum dynamics of non-interacting fermionic systems; and establishes clear rules for determining whether a static or dynamical fermionic problem can be solved in terms of bosonic variables and how to articulate the fundamental expressions to be solved in the bosonic representation. This lays the foundation for the controlled application of such maps to more complex systems, such as those coupled to nuclear motions or exhibiting correlation effects, with the more complete quantum-classical hierarchy.

\section{Fermion to boson mapping}
\label{sec:quadratic-fermionic-hamiltonian}

The basic assumption we want to test is conditions under which it is  valid to make the following substitution,
\begin{subequations}\label{eq:fermion-to-boson-inexact-map}
\begin{align}
    \hat{c}^{\dagger}_{j} \ \ ``\mapsto" \ \ \hat{b}^{\dagger}_{j}, \label{eq:one-fermion-to-one-boson-creation}\\
    \hat{c}_{j} \ \ ``\mapsto"  \ \ \hat{b}_{j} \label{eq:one-fermion-to-one-boson-annihilation}.
\end{align}
\end{subequations}
Clearly, in most cases this does not constitute an exact map since bosonic and fermionic operators follow fundamentally different algebras, i.e., fermionic operators anticommute, while bosonic ones commute, 
\begin{subequations}
\begin{align}
    [\hat{c}_j, \hat{c}_k^{\dagger}]_{+} &\equiv \hat{c}_j\hat{c}_k^{\dagger} + \hat{c}_k^{\dagger}\hat{c}_j = \delta_{j,k}, \label{eq:fermionic-anticommutation-relation}\\
    [\hat{b}_j, \hat{b}_k^{\dagger}]_{-} &\equiv \hat{b}_j\hat{b}_k^{\dagger} - \hat{b}_k^{\dagger}\hat{b}_j = \delta_{j,k}. \label{eq:bosonic-commutation-relation}
\end{align}
\end{subequations}
Here we show when this inexact map can exactly capture the static and dynamical observables of systems described by quadratic fermionic Hamiltonians of the form, 
\begin{equation} \label{eq:general-quadratic-fermion-hamiltonian}
    H = \sum_{j,k} h_{j,k} \hat{c}_{j}^{\dagger}\hat{c}_{k},
\end{equation}
where $h_{j,k}$ are the matrix elements of the single-particle Hamiltonian. Such Hamiltonians form the basis of the description of a variety of phenomena, ranging from elastic charge transport in crystals \cite{holsteinStudiesPolaronMotion1959a, troisiChargeTransportRegimeCrystalline2006, fratiniTransientLocalizationScenario2016} and through nanojunctions \cite{Evers2019, CohenGalperin2020} to catalytic activity at electrochemical interfaces \cite{chidseyFreeEnergyTemperature1991, sehCombiningTheoryExperiment2017, warburtonTheoreticalModelingElectrochemical2022, santosModelsElectronTransfer2022}. However, while such Hamiltonians form the basis of mean-field treatments of correlation effects, these cannot account for significant correlation effects, such as the Kondo effect \cite{Hewson1993-Kondo-book}, Coulomb blockades \cite{averinCoulombBlockadeSingleelectron1986, beenakkerTheoryCoulombblockadeOscillations1991, Levy2019}, or superconductivity \cite{Auerbach1998, dagottoCorrelatedElectronsHightemperature1994, orensteinAdvancesPhysicsHighTemperature2000, leeDopingMottInsulator2006}. Despite the inability of quadratic Hamiltonians to account for these more exotic effects, showing when the bosonic representation in Eq.~(\ref{eq:fermion-to-boson-inexact-map}), and therefore any quantum-classical theories derived based on it, can capture the statics and dynamics of these systems represents an essential first step in the development of more advanced, even if approximate, treatments that can account for additional complexity, such as coupling to nuclear motions and correlation effects. 

Before detailing how the map in Eqs.~(\ref{eq:one-fermion-to-one-boson-creation}) and (\ref{eq:one-fermion-to-one-boson-annihilation}) can be used to calculate static and dynamical observables, we first summarize a few formal results about quadratic fermionic Hamiltonians and set the notation for our derivations. In particular, access to the unitary transformation, $\mathbf{U}$, that diagonalizes the single-particle Hamiltonian, $\mathbf{h} \rightarrow \mathbf{U}^{-1}\mathbf{h}\mathbf{U} = \mathbf{E}$, where $E_{j,k} = \varepsilon_j \delta_{j,k}$, allows one to exactly calculate any dynamical property of the system. This is because $\mathbf{U}$ allows one to reexpress Eq.~(\ref{eq:general-quadratic-fermion-hamiltonian}) as a collection of noninteracting fermions, $H = \sum_{j} \varepsilon_{j} \hat{C}_{j}^{\dagger}\hat{C}_{k}$, where the new fermionic creation and annihilation operators can be expressed as a linear combination of the old fermionic operators, $\hat{C}_{j} = \sum_{j,k} U^{-1}_{j,k}\hat{c}_{k}$ and $\hat{C}^{\dagger}_{j} = \sum_{k} U_{k,j}\hat{c}^{\dagger}_{k}$. The time-dependence of the noninteracting fermions, $\hat{C}(t) = e^{-i\varepsilon_k t}\hat{C}$, can then be used to construct the time-dependence of the original fermions, 
\begin{subequations}\label{eq:fermions-evolution}
\begin{align}
    \hat{c}_j^{\dagger}(t) &= \sum_{m} G^{*}_{m,j}(t) \hat{c}_m^{\dagger}, \label{eq:fermion-creation-evolution}\\
    \hat{c}_j(t) &= \sum_{m} G_{j,m}(t) \hat{c}_m, \label{eq:fermion-annihilation-evolution}
\end{align}
\end{subequations}
where and the time-dependence of an operator is given by the Heisenberg picture, $\mathcal{O}(t) = e^{iHt}\mathcal{O}(t)e^{-iHt}$, and the time-dependent coefficients take the form
\begin{equation}\label{eq:fermionic-G-time-dependent-matrix}
    G_{n,m} = \sum_{r} U_{n,r} e^{-i\varepsilon_r t}  U^{-1}_{r,m}.
\end{equation}
Because any operator can be written as a linear combination of products of individual creation and annihilation operators, Eqs.~(\ref{eq:fermion-creation-evolution}), (\ref{eq:fermion-annihilation-evolution}), and (\ref{eq:fermionic-G-time-dependent-matrix}) can be used to construct the time dependence of arbitrary operators, thus allowing one to evaluate any dynamical observable in the original representation.

Applying the map in Eqs.~(\ref{eq:fermion-to-boson-inexact-map}) to the quadratic fermionic Hamiltonian in Eq.~(\ref{eq:general-quadratic-fermion-hamiltonian}) yields, 
\begin{equation}\label{eq:general-quadratic-fermion-hamiltonian-in-bosonic-operators}
\begin{split}
    H \mapsto \mathcal{H} = \sum_{j,k} h_{j,k} \hat{b}_{j}^{\dagger}\hat{b}_{k},
\end{split}
\end{equation}
where $\{ \hat{b}_j^{\dagger}, \hat{b}_j \}$ are bosonic creation and annihilation operators. Subject to this Hamiltonian, the time evolution of the bosonic operators is given by the same time-dependent coefficients as in the fermionic case, 
\begin{subequations} \label{eq:bosons-evolution}
\begin{align}
    \hat{b}_j^{\dagger}(t) &= \sum_{m} G^{*}_{m,j}(t) \hat{b}_m^{\dagger}, \label{eq:boson-creation-evolution}\\
    \hat{b}_j(t) &= \sum_{m} G_{j,m}(t) \hat{b}_m, \label{eq:boson-annihilation-evolution}
\end{align}
\end{subequations}
and $G_{n,m}(t)$ is given by Eq.~(\ref{eq:fermionic-G-time-dependent-matrix}). 

From Eqs.~(\ref{eq:fermions-evolution}), (\ref{eq:fermionic-G-time-dependent-matrix}), and (\ref{eq:bosons-evolution}) it can be seen that the bosonic representation in Eq.~(\ref{eq:fermion-to-boson-inexact-map}) provides an exact means to calculate the time-dependence of any mapped fermionic operator. However, just observing that the bosonic representation captures the exact time-dependence of fermionic operators \textit{is not sufficient} to actually evaluate their matrix elements. It is also necessary to have a well defined basis with respect to which one can calculate matrix elements.

To define an appropriate basis for the bosonic representation that allows one to evaluate the matrix elements of static and dynamic observables of fermionic operators, it is necessary to have a prescription that maps the fermionic basis to a physically restricted bosonic one. Specifically, due to the anticommutivity of fermions, Eq.~(\ref{eq:fermionic-anticommutation-relation}), their Hilbert space consists of only two states per fermionic mode. In contrast, the commutivity of bosons, Eq.~(\ref{eq:bosonic-commutation-relation}), results in an infinite dimensional Hilbert space for every bosonic mode. To connect the sizes of these two Hilbert spaces, one can restrict the size of the bosonic Hilbert space to match that of fermions. While there are several ways of enforcing this restricted Hilbert space \cite{Schwinger1965, Holstein1940, Auerbach1998, Montoya2018}, here we focus on one which simply truncates the Hilbert space of each bosonic mode two its unoccupied and singly occupied states. Such a truncation naturally arises in lattice models where bosons on the same site repel strongly, i.e., a Bose-Hubbard model \cite{fisherBosonLocalizationSuperfluidinsulator1989} in the limit of $U \rightarrow \infty$, where $U$ is the on-site two-boson interaction. In these cases, the Hilbert space of the $j$th bosonic mode contains only two states,
\begin{equation}\label{eq:restricted-boson-basis}
    \{ \ket{n_j} \} \equiv \{ \ket{0_j}, \ket{1_j} \},
\end{equation}
where $n$ indicates the occupation of the $k$th mode, which exactly matches the Hilbert space of the $k$th fermionic mode. Bosons whose Hilbert space spans only the unoccupied and first occupied states are called hard-core bosons. As in the fermionic case, for a system consisting of $M$ hard-bosonic modes, the many-body Fock state, $\ket{\mathbf{n}} = \ket{n_1, n_2, ..., n_M}$ is determined by the occupation number of each mode. Now with commensurate Hilbert spaces, the main difference between the fermionic and bosonic cases lies in the statistics of the operators which, upon acting on the many-body state, can lead to occupation dependent phases. Specifically, fermionic anticommutivity implies that, 
\begin{subequations}
\begin{align}
    \hat{c}_j^{\dagger} \ket{n_1 ... n_j ... n_M} &= \delta_{n_j,1} (-1)^{h(j)} \ket{n_1 ... n_j+1 ... n_M},\\
    \hat{c}_j \ket{n_1 ... n_j ... n_M} &= \delta_{n_j,0} (-1)^{h(j)} \ket{n_1 ... n_j-1 ... n_M},
\end{align}
\end{subequations}
where 
\begin{equation}\label{eq:counter-function-for-phase}
    h(j|k) = \sum_{\substack{l=1\\l \neq k}}^{j-1} n_l 
\end{equation}
is a counter function that accounts for the exponent of the negative phase acquired when the $j$th creation or annihilation operator acts on a many-body state containing excitations in indices $l < j$. For cases where additional indices are specified after the vertical line, e.g., $k$ in Eq.~(\ref{eq:counter-function-for-phase}), the sum excludes these indices. In contrast, the commutative nature of bosonic operators does not lead to these phases,
\begin{subequations}
\begin{align}
    \hat{b}_j^{\dagger} \ket{n_1 ... n_j ... n_M} &= \delta_{n_j,1} \ket{n_1 ... n_j+1 ... n_M},\\
    \hat{b}_j \ket{n_1 ... n_j ... n_M} &= \delta_{n_j,0} \ket{n_1 ... n_j-1 ... n_M}.
\end{align}
\end{subequations}

As we discuss below, there is a subset of problems where the bosonic representation in Eq.~(\ref{eq:fermion-to-boson-inexact-map}) coupled with the hard-core truncation of the Hilbert space in Eq.~(\ref{eq:restricted-boson-basis}) allows one to correctly calculate certain matrix elements of static and time-dependent fermionic operators. The bosonic representation leads to difficulties only when occupation-dependent phases arise in the fermionic representation. Thus, here we provide a simple guide for when the hard-core bosonic representation allows one to correctly capture the matrix elements and correlation functions of fermionic systems subject to a quadratic Hamiltonian of the form given by Eq.~(\ref{eq:general-quadratic-fermion-hamiltonian}).

Before turning to this analysis, we briefly note the relation of the map in Eq.~(\ref{eq:fermion-to-boson-inexact-map}) to other techniques used previously, albeit in generally different contexts. We note, for instance, that the replacement of fermionic creation and annihilation operators examined here is distinctly different from bosonization \cite{Giamarchi2004, senechalTheoreticalMethodsStrongly}, a technique that has been used with great success to interrogate the dynamics of fermionic problems and spin-chains in one dimension. Bosonization is based on the insight that particle-hole excitations are bosonic in character. As such, it is the density fluctuations, not the individual fermionic creation and annihilation operators, that are mapped to bosonic creation and annihilation operators \cite{Giamarchi2004, Senechal2004-Theoretical-Methods-for-Strongly-Correlated-Electrons}. In contrast, the approach considered here directly replaces fermionic creation and annihilation operators by bosonic ones.

The map in Eq.~(\ref{eq:fermion-to-boson-inexact-map}) also differs from maps based on multilevel systems, which often reduce to adopting the commonly used independent electron approximation \cite{Kornyshev1985, Sebastian1989, Norsvok1990, Kondov2007}. This approximation replaces creation-annihilation pair products with their single-particle orbitals, 
\begin{equation}\label{eq:iea-creation-annihilation-pair-map}
    \hat{c}_{j}^{\dagger}\hat{c}_k \ \  ``\mapsto" \ \  \ket{j}\bra{k}. 
\end{equation}
Once expressed in this form, one can then apply the Meyer-Miller-Stock-Thoss transformation \cite{Meyer1979, Stock1997}, which maps outer products of the form in Eq.~(\ref{eq:iea-creation-annihilation-pair-map}) to bosonic variables. Indeed, following this procedure, one would again obtain the mapped Hamiltonian in Eq.~(\ref{eq:general-quadratic-fermion-hamiltonian-in-bosonic-operators}). However, the $M$-dimensional physical basis for this problem spans the singly occupied bosonic Fock states, where the $k$th vector consists of the product of the singly occupied state of the $k$th boson and the ground state of all other modes. As we show in Appendix \ref{app:independent-electron-approximation}, this approximation only allows one to get, at most, the expectation value single-time correlation functions of one-body operators. In contrast, as we demonstrate below, the fermion to boson map in Eq.~(\ref{eq:fermion-to-boson-inexact-map}) permits for the calculation of single and multi-time correlation functions of up to two-body operators. 

Finally, we remark that, once a problem is expressed in terms of bosonic creation and annihilation operators, one can employ the Cartesian coordinate representation of these operators, i.e., 
\begin{subequations}
\begin{align}
    \hat{b}^{\dagger}_{j} &= (\hat{q} - i\hat{p})/ \sqrt{2},\\
    \hat{b}_{j} &= (\hat{q} + i\hat{p})/ \sqrt{2},
\end{align}
\end{subequations}
to express the mapped observables and Hamiltonian in terms of Cartesian phase space operators, 
\begin{equation}\label{eq:general-quadratic-fermion-hamiltonian-in-cartesian-operators}
\begin{split}
    \mathcal{H} &= \frac{1}{2} \sum_{j, k} h_{j,k} \Big[\hat{q}_j\hat{q}_k + \hat{p}_j\hat{p}_k - \delta_{j,k} + i \big(\hat{q}_j\hat{p}_k - \hat{p}_j\hat{q}_k \big) \Big],
\end{split}
\end{equation}
A particularly advantageous property of quadratic Hamiltonians of the form in Eq.~(\ref{eq:general-quadratic-fermion-hamiltonian-in-cartesian-operators}) is that the classical dynamics of the resulting Cartesian variables captures the exact quantum dynamics of such systems \cite{Stock1997, Thoss1999, sunBosonicPerspectiveClassical2021}, allowing one to calculate quantum correlation functions at the cost of classical calculations. When considering more complex situations, such as those where fermionic degrees or freedom are coupled to nuclear motions, the phase space formulation provides a convenient starting place for quantum-classical approximations. This compatibility with the quantum-classical hierarchy \cite{Miller2001a, Stock2005, Kapral2015, Lee2016b, Crespo-Otero2018,  Bonnet2020}, which ranges from efficient but generally inaccurate mean-field approaches \cite{McLachlan1964, Stock1995, Sun1997a, Shi2004, Cotton2013} to more accurate but resource intensive approaches \cite{Bonella2005, Kim2008, Huo2012, Hsieh2012, Kapral2015}, is particularly compelling, as one can treat the effect of an external environment, such as nuclear motions, on the same theoretical footing. In addition, as recent work has demonstrated \cite{Shi2004a, Kelly2013, Kelly2015, Pfalzgraff2015, Montoya2016a, KellyMontoya2016, Montoya2017, Pfalzgraff2019, Mulvihill2019, Mulvihill2019a, Mulvihill2021, Mulvihill2021b, Mulvihill2022}, quantum-classical methods can be successfully combined with the generalized quantum master equation framework to improve the efficiency and accuracy of quantum-classical schemes.  Thus, elucidating when the map in Eq.~(\ref{eq:fermion-to-boson-inexact-map}) can be exploited to exactly calculate observables in many-fermion problems can set the stage for the controlled extension and application of such a map to more complex systems of interest where an exact solution is difficult to obtain.

\subsection{Matrix elements of static operators}
\label{ssec:matrix-elements-of-static-operators}

Perhaps one of the simplest and most fundamental questions one must ask about the the feasibility of replacing fermionic by bosonic operators according to Eq.~(\ref{eq:fermion-to-boson-inexact-map}) concerns the criteria that an arbitrary operator needs to satisfy for its matrix elements both in the fermionic and bosonic representations to be equivalent. Unfortunately, the matrix elements of most fermionic operators contain occupation-dependent phases which the bosonic representation does not capture. In contrast, as we show below, the class of fermionic operators for which the matrix elements are equivalent in either representation is limited.

We begin this discussion by specifying the fermionic operators for which one may expect matrix elements to contain occupation-dependent phases. For example, consider the matrix elements of a single creation (or annihilation) operator for the $j$th mode,
\begin{equation} \label{eq:definition-occupation-dependent-phase}
    \bra{\mathbf{n}} \hat{c}_j^{\dagger} \ket{\mathbf{n}'} = (-1)^{h(j)} \delta_{n_j', n_j+1} F(\mathbf{n}, \mathbf{n}'|j),
\end{equation}
where $\mathbf{n} = (n_1, n_2, ..., n_m)$ and $\mathbf{n}' = (n_1', n_2', ..., n_M')$ denote two arbitrary sets of occupation numbers which determine the Fock states used to obtain the value of all matrix elements, and \begin{equation}\label{eq:definition-contraction-many-body-basis-delta-functions}
    F(\mathbf{n}, \mathbf{n}'| \mathbf{j} ) \equiv \prod_{\substack{k=1\\ k \neq \{ j \}}} \delta_{n_k, n_k'}
\end{equation}
enforces the equal occupation of the modes across the $\mathbf{n}$ and $\mathbf{n}'$ sets, with the exception of excluded indices $\mathbf{j}$. The matrix element in Eq.~(\ref{eq:definition-occupation-dependent-phase}) contains an occupation-dependent phase, $(-1)^{\sum_{l=1}^{j-1} n_l}$, which would not arise if one evaluated the matrix element in the bosonic representation,
\begin{equation} \label{eq:definition-occupation-dependent-phase-lack-for-bosons}
    \bra{\mathbf{n}} \hat{b}_j^{\dagger} \ket{\mathbf{n}'} = \delta_{n_j', n_j+1} F(\mathbf{n}, \mathbf{n}'|j).
\end{equation}
Previous fermion mapping approaches \cite{Miller1986, Montoya2018} that explicitly account for fermionic anticommutivity contain the appropriate nonlocal factors that account for the phase. Despite not recovering the correct value of matrix elements with a finite weight, it is noteworthy matrix elements with zero weight are correctly captured in the bosonic representation. It is straightforward to extend this conclusion to any fermionic operator containing a lone creation or annihilation operator corresponding to an arbitrary mode $j$. Instead, the only way to reliably remove these occupation-dependent phases is to restrict one's attention to operators consisting of \textit{products of single mode creation-annihilation pairs}. However, even within this family, we must place certain restrictions on the way these products are ordered. 

To understand the restrictions that one must place on operators consisting of products of single mode creation-annihilation pairs, it is essential to consider the orderings of operators that do not lead to occupation-dependent phases. Perhaps the simplest type of operator that allows for the evaluation of its matrix elements in the bosonic representation is one which is \textit{pairwise-ordered}. Pairwise-ordered operators are those where single mode creation-annihilation pairs appear next to each other, e.g., $\hat{c}_j\hat{c}_j^{\dagger}\hat{c}_k^{\dagger}\hat{c}_k$, $\hat{c}_j^{\dagger}\hat{c}_j\hat{c}_k\hat{c}_k^{\dagger}$, $\hat{c}_j\hat{c}_j^{\dagger}\hat{c}_k\hat{c}_k^{\dagger}$, where $j \neq k$. In fact, \textit{one can exactly evaluate the matrix elements of any ordering of products of single mode creation-annihilation pairs in the bosonic representation only if the operator product requires at most an even number of permutations to achieve a pairwise ordered form}. The proof of this statement is provided in Appendix~\ref{sec:proof-pairwise-ordered-operators-static-matrix-elements}. These operators include \textit{normal} and \textit{anti-normal ordered} operators, such as $\hat{c}_j^{\dagger}\hat{c}_k^{\dagger}\hat{c}_k\hat{c}_j$ and $\hat{c}_j\hat{c}_k\hat{c}_k^{\dagger}\hat{c}_j^{\dagger}$, respectively, where $j < k$. Normal-ordered products are those where all creation operators appear to the left in increasing order from left to right, while annihilation operators appear to the right, in decreasing order from left to right. 

Because of their central importance, we call orderings of products of single mode creation-annihilation pairs \textit{proper-ordered} if their matrix elements can be it calculated in the bosonic representation, i.e., if it takes an even number of permutations to rearrange the operators into a pairwise ordered form. An important consequence of this is that one can also use the bosonic representation to evaluate the matrix elements of any power of proper-ordered operators (see Appendix \ref{sec:proof-pairwise-ordered-operators-static-matrix-elements}), and therefore functions of such operators. 

In summary, using the inexact bosonic map of Eq.~(\ref{eq:fermion-to-boson-inexact-map}), it is possible to exactly recover (i) the diagonal and off-diagonal matrix elements of products containing odd numbers of single-mode creation or annihilation operators as long as these are equal to zero and (ii) \textit{all} matrix elements of proper-ordered products of single mode creation-annihilation pairs and their functions. Hence, the only nonzero elements that one can capture in the bosonic representation are the diagonal matrix elements of proper-ordered operators. In contrast, the nonzero matrix elements of operators containing products of odd numbers of creation or annihilation operators and products of single mode creation-annihilation pairs are not proper-ordered will generally contain occupation-dependent phases, which are not captured in the bosonic representation.

\subsection{Matrix elements of time-dependent operators}
\label{ssec:matrix-elements-of-time-dependent-operators}

Guided by the fact that the time-dependence of fermionic is equivalent to that of bosonic operators subject to an analogous Hamiltonian, one might imagine that, as in the case of static operators, it should be possible to calculate the diagonal matrix elements of all time-evolved proper-ordered operators in the bosonic representation. However, evolving a proper-ordered operator results in a linear combination of operators with time-dependent coefficients, some of which may not be proper-ordered. 

For example, consider the time evolution of a single product of creation-annihilation pair, 
\begin{equation} \label{eq:evolved-fermionic-creation-annihilation-pair}
    \hat{c}_j(t)\hat{c}_j^{\dagger}(t) = \sum_{p,q} G^{*}_{p,j}(t)G_{j,q}(t)    \hat{c}_q\hat{c}_p^{\dagger},
\end{equation}
where the time-dependence of the creation and annihilation operators is given by Eqs.~(\ref{eq:fermion-creation-evolution}) and (\ref{eq:fermion-annihilation-evolution}), respectively. We can then split this sum into two contributions, one coming from cases where $p=q$ and one from cases where $p \neq q$, 
\begin{subequations}
\begin{align}
    \bra{ \mathbf{n}}  \hat{c}_p\hat{c}_p^{\dagger} \ket{ \mathbf{n}'} &= F(\mathbf{n}, \mathbf{n}') \delta_{n_p, 0},\label{eq:evolved-fermionic-creation-annihilation-pair-diagonal}\\
    \bra{ \mathbf{n}} \hat{c}_q\hat{c}_p^{\dagger}\ket{ \mathbf{n}'} &= F(\mathbf{n}, \mathbf{n}'| p, q) \delta_{n'_{p}+1, n_p} \delta_{n'_{q}, n_q+1} (-1)^{h(p) + h(q|p)}.\label{eq:evolved-fermionic-creation-annihilation-pair-offdiagonal}
\end{align}
\end{subequations}
The case where $p=q$ in Eq.~(\ref{eq:evolved-fermionic-creation-annihilation-pair-diagonal}) corresponds to a proper-ordered static operator whose matrix elements can be captured in the bosonic representation. In contrast, the case where $p \neq q$ in Eq.~(\ref{eq:evolved-fermionic-creation-annihilation-pair-offdiagonal}) is not proper-ordered, leads to occupation-dependent phases on its finite off-diagonal matrix elements, and can therefore not be captured by the bosonic representation. However, the bosonic representation is able to capture the diagonal matrix elements of the operator in this latter case, which are equal to zero. In other words, the bosonic representation is able to capture the diagonal matrix elements of arbitrary creation-annihilation pairs correctly, which implies that it captures the diagonal matrix elements of time-evolved creation-annihilation pairs whether these arise from the same or different modes,
\begin{subequations}\label{eq:diagonal-matrix-element-general-time-dependent-1-creation-annihilation-pair}
\begin{align}
    \bra{ \mathbf{n}}\hat{c}^{\dagger}_j(t_1)\hat{c}_{k}(t_2)\ket{ \mathbf{n}} &= \sum_{j', k'} G_{j',j}^{*}(t_1)G_{k,k'}(t_2) \delta_{j',k'} \delta_{n_k', 1} \nonumber \\
    &= \bra{\mathbf{n}}\hat{b}^{\dagger}_j(t_1)\hat{b}_{k}(t_2)\ket{ \mathbf{n}} ,\\
    \bra{ \mathbf{n}}\hat{c}_{k}(t_2)\hat{c}^{\dagger}_j(t_1)\ket{ \mathbf{n}} &= \sum_{j',k'} G_{j',j}^{*}(t_1)G_{k,k'}(t_2) \delta_{j',k'}\delta_{n_k', 0} \nonumber \\
    &= \bra{ \mathbf{n}}\hat{b}_{k}(t_2)\hat{b}^{\dagger}_j(t_1)\ket{ \mathbf{n}}.
\end{align}
\end{subequations}
This conclusion is consistent with the statement in the previous section which states that one can recover the both the nonzero matrix elements of proper-ordered products of time-independent operators and the zero matrix elements of products containing an odd number of single-mode creation and/or annihilation operators.  

One may then ask whether the time-dependent operator of interest can take a more complex form. In the following, we demonstrate that the most complicated form an operator can have for which one can still recover the diagonal matrix elements correctly in the bosonic representation is one containing at most a quartic product of two creation-annihilation pairs. This product can be ordered in at most two configurations,  
\begin{subequations}\label{eq:diagonal-matrix-element-general-quartic-product-2-creation-annihilation-pairs}
\begin{align}
    \bra{ \mathbf{n}}\hat{c}^{\dagger}_j\hat{c}_{k}\hat{c}_l^{\dagger}\hat{c}_{m}\ket{ \mathbf{n}} &= \Big(\delta_{j,k}\delta_{l,m} \delta_{n_k, 1} + (1 - \delta_{k,m})\delta_{j,m} \delta_{k,l}  \delta_{n_k, 0}\Big)\delta_{n_m, 1} \nonumber \\
    &= \bra{ \mathbf{n}}\hat{b}^{\dagger}_j\hat{b}_{k}\hat{b}_l^{\dagger}\hat{b}_{m}\ket{ \mathbf{n}} , \label{eq:diagonal-matrix-element-general-quartic-product-2-creation-annihilation-pairs-creation-annihilation}\\
    \bra{ \mathbf{n}}\hat{c}_{k}\hat{c}^{\dagger}_j\hat{c}_{m}\hat{c}_l^{\dagger}\ket{ \mathbf{n}} &= \Big(\delta_{j,k}\delta_{l,m} \delta_{n_m, 0} + (1 - \delta_{k,m})\delta_{j,m} \delta_{k,l} \delta_{n_m, 1}\Big)\delta_{n_k, 0} \nonumber \\
    &= \bra{ \mathbf{n}}\hat{b}_{k}\hat{b}^{\dagger}_j\hat{b}_{m}\hat{b}_l^{\dagger}\ket{ \mathbf{n}},\label{eq:diagonal-matrix-element-general-quartic-product-2-creation-annihilation-pairs-annihilation-creation}
\end{align}
\end{subequations}
where the order of the first creation-annihilation (Eq.~(\ref{eq:diagonal-matrix-element-general-quartic-product-2-creation-annihilation-pairs-creation-annihilation})) or annihilation-creation (Eq.~(\ref{eq:diagonal-matrix-element-general-quartic-product-2-creation-annihilation-pairs-annihilation-creation})) sequence determines the sequence of the second pair. These products can arise from the time evolved versions of the following two types of operators, 
\begin{widetext}
\begin{subequations}\label{eq:diagonal-matrix-element-general-time-dependent-2-creation-annihilation-pairs}
\begin{align}
    \bra{ \mathbf{n}}\hat{c}^{\dagger}_{j}(t_1)\hat{c}_{k}(t_2)\hat{c}_{l}^{\dagger}(t_3)\hat{c}_{m}(t_4)\ket{ \mathbf{n}} &= \sum_{j',k',l',m'} \mathcal{G}_{j,k,l,m}^{j',k',l',m'}(t_1, t_2, t_3, t_4) \bra{ \mathbf{n}}\hat{c}^{\dagger}_{j'}\hat{c}_{k'}\hat{c}_{l'}^{\dagger}\hat{c}_{m'}\ket{ \mathbf{n}} = \bra{ \mathbf{n}}\hat{b}^{\dagger}_{j}(t_1)\hat{b}_{k}(t_2)\hat{b}_{l}^{\dagger}(t_3)\hat{b}_{m}(t_4)\ket{ \mathbf{n}},\\
    \bra{ \mathbf{n}}\hat{c}_{k}(t_2)\hat{c}^{\dagger}_{j}(t_1)\hat{c}_{m}(t_4)\hat{c}_{l}^{\dagger}(t_3)\ket{ \mathbf{n}} &= \sum_{j',k',l',m'} \mathcal{G}_{j,k,l,m}^{j',k',l',m'}(t_1, t_2, t_3, t_4) \bra{ \mathbf{n}}\hat{c}_{k'}\hat{c}^{\dagger}_{j'}\hat{c}_{m'}\hat{c}_{l'}^{\dagger}\ket{ \mathbf{n}} = \bra{ \mathbf{n}}\hat{b}_{k}(t_2)\hat{b}^{\dagger}_{j}(t_1)\hat{b}_{m}(t_4)\hat{b}_{l}^{\dagger}(t_3)\ket{ \mathbf{n}},
\end{align}
\end{subequations}
\end{widetext}
where
\begin{equation}
    \mathcal{G}_{j,k,l,m}^{\alpha, \beta, \gamma, \delta}(t_1, t_2, t_3, t_4) = G^{*}_{j',j}(t_1) G_{k,k'}(t_2) G^{*}_{l',l}(t_3) G_{m,m'}(t_4).
\end{equation}
We have used different time indices, $\{t_1, t_2, t_3, t_4\}$, to emphasize these operators can indeed be evolved to different time, yielding the same sum over quartic products of creation-annihilation pairs in Eqs.~(\ref{eq:diagonal-matrix-element-general-quartic-product-2-creation-annihilation-pairs}), albeit with different time-dependent coefficients. 

If one tries to go beyond the product of two creation-annihilation pairs of the form given in Eqs.~(\ref{eq:diagonal-matrix-element-general-quartic-product-2-creation-annihilation-pairs}), the diagonal matrix elements start to contain occupation-dependent phases and therefore cannot be reliably captured by the bosonic representation. For instance, consider the matrix elements of the three creation-annihilation pair product, 
\begin{equation}\label{eq:fermionic-three-pair-expectation}
    \bra{ \mathbf{n}}\hat{c}^{\dagger}_{j}\hat{c}_{k}\hat{c}_{l}^{\dagger}\hat{c}_{m}\hat{c}_{p}^{\dagger}\hat{c}_{q}\ket{ \mathbf{n}},
\end{equation}
where the operator indices can take any value in $\{1, ..., M \}$. In this case, one of the distinct pair contractions of the indices that contributes to the diagonal matrix elements, i.e., $j=m$, $k=p$, and $l=q$, where $j \neq k \neq l$, leads to an occupation-dependent phase, 
\begin{equation}
\begin{split}
    \bra{ \mathbf{n}}\hat{c}^{\dagger}_{j}\hat{c}_{k}\hat{c}_{l}^{\dagger}\hat{c}_{j}\hat{c}_{k}^{\dagger}\hat{c}_{l}\ket{ \mathbf{n}} &= (-1)^{3} \delta_{n_j,1}\delta_{n_k,0} \delta_{n_l,0} \\
    &\neq \bra{ \mathbf{n}}\hat{b}^{\dagger}_{j}\hat{b}_{k}\hat{b}_{l}^{\dagger}\hat{b}_{j}\hat{b}_{k}^{\dagger}\hat{b}_{l}\ket{ \mathbf{n}}.
\end{split}
\end{equation}
As the operator products increase in complexity, i.e., consist of a larger number of creation-annihilation pairs, the number of contributions that contain negative phases becomes more significant. This finding implies that operators of the form given in Eq.~(\ref{eq:diagonal-matrix-element-general-time-dependent-2-creation-annihilation-pairs}) are the most complex operators for which one can recover diagonal matrix elements correctly in the bosonic representation. Furthermore, because expansion of a function in its Taylor series leads to a sum over different powers of its arguments, the bosonic representation is generally unable to correctly capture the matrix elements of functions of operators, except in special cases where the expansion can be truncated at a low order where the resulting operators conform to the criterion of Eqs.~(\ref{eq:diagonal-matrix-element-general-time-dependent-1-creation-annihilation-pair}) and (\ref{eq:diagonal-matrix-element-general-time-dependent-2-creation-annihilation-pairs}). Since the bosonic representation is able to correctly capture only the diagonal matrix elements of operators consisting of one or two creation-annihilation pairs, such as those given by Eqs.~(\ref{eq:diagonal-matrix-element-general-time-dependent-1-creation-annihilation-pair}) and (\ref{eq:diagonal-matrix-element-general-time-dependent-2-creation-annihilation-pairs}), we refer to these operators and their allowed orderings as time-proper-ordered.

\subsection{Observables and initial conditions}
\label{ssec:boson-version-observables-and-initial-conditions}

In Secs.~\ref{ssec:matrix-elements-of-static-operators} and \ref{ssec:matrix-elements-of-time-dependent-operators}, we showed which matrix elements of static and dynamical operators can be obtained via the bosonic representation in Eq.~(\ref{eq:fermion-to-boson-inexact-map}). In this section, we exploit these insights to determine the types of initial conditions and observables for which one can \textit{exactly} calculate nonequilibrium averages and general time-correlation functions when using the transformation in Eq.~(\ref{eq:fermion-to-boson-inexact-map}).

We begin by considering a system whose many-body Hilbert space is constructed using $M$ single-particle states and is thus contains $2^M$ many-body states. As done previously, we restrict our attention to Hamiltonians of the form given by Eq.~(\ref{eq:general-quadratic-fermion-hamiltonian}). We are interested in considering the calculation of time-dependent averages such as correlation functions of the form, 
\begin{equation}\label{eq:n-time-correlation-function}
\begin{split}
    \langle \hat{B}(\mathbf{t}) \rangle &= \mathrm{Tr}[\hat{\rho} \hat{B}(\mathbf{t})],
\end{split}
\end{equation}
where $\hat{\rho}$ encodes the initial condition of the system, $\hat{B}(\mathbf{t}) = \hat{B}_j(0)\hat{B}_k(t_1) ... \hat{B}_l(t_N)$ consists of a product of operators $\{ \hat{B}_x \}$ corresponding to individual creation or annihilation operators and $\mathbf{t} = (t_0, t_1, ..., t_N)$ is the set of time arguments of each operator. We have maintained distinct indices for the time arguments of these operators to emphasize that there is no restriction on the identity of these arguments.

Equation~(\ref{eq:n-time-correlation-function}) encompasses a wide range of observables and physical situations. For example, $\hat{\rho}$ can correspond to a nonequilibrium state normally associated with charge transport setups where the leads are in local thermal and chemical equilibrium while the impurity is either occupied or unoccupied, or an equilibrium one where $\hat{\rho}$ assumes the form of the canonical density of the entire system. The operators $\hat{B}_j$ can take the form of a population, e.g., of the impurity or reservoir level(s), current, or the unit operator, $\hat{\mathbf{1}}$. As such, by elucidating the criteria that Eq.~(\ref{eq:n-time-correlation-function}) needs to satisfy, we also determine the restrictions on the nonequilibrium averages and time correlation functions that one may calculate in the bosonic representation.

As we demonstrated in Secs.~\ref{ssec:matrix-elements-of-static-operators} and \ref{ssec:matrix-elements-of-time-dependent-operators}, the only nonzero elements that are correctly captured for any operator, static or time-dependent, are the diagonal matrix elements of a subset of operators. Specifically, in the case of static operators, only proper-ordered products of single mode creation-annihilation pairs allows for the replacement of fermionic by bosonic operators with a limited Hilbert space, while for time-dependent operators, these need to conform to a more stringent standard, i.e., proper-time-ordered operators consisting of, at most, two products of creation-annihilation pairs. Hence, diagonal proper-ordered initial conditions, $\rho$, would allow one to reexpress the trace operation as a sum over the diagonal matrix elements in the basis of all Fock states,
\begin{equation}\label{eq:n-time-correlation-function-diagonal-rho}
\begin{split}
    \langle \hat{B}(\mathbf{t}) \rangle &= \sum_{ \{ \mathbf{n} \}}\bra{\mathbf{n}} \hat{\rho} \ket{\mathbf{n}} \bra{\mathbf{n}}\hat{B}(\mathbf{t})\ket{\mathbf{n}}.
\end{split}
\end{equation}
In general, these initial conditions take the form
\begin{equation}\label{eq:diagonal-rho}
    \rho = \prod_{m=1}^{M} \rho_m(\hat{c}_m^{\dagger} \hat{c}_m),
\end{equation}
where
\begin{equation}\label{eq:diagonal-initial-condition-indivdual-fermion}
    \rho_m(\hat{c}_m^{\dagger}\hat{c}_m) = (1 - p_m) \hat{c}_m\hat{c}_m^{\dagger} + p_m\hat{c}_m^{\dagger}\hat{c}_m,
\end{equation}
and $p_m = \mathrm{Tr}[\rho \hat{c}_{m}^{\dagger}\hat{c}_{m}]$. The form for $\rho$ in Eq.~(\ref{eq:diagonal-rho}) encompasses widely used initial conditions, such as that corresponding to an nonequilibrium state of an impurity in its occupied ($\rho_m = \hat{c}^{
\dagger}_m\hat{c}_m$) or unoccupied ($\rho_m = \hat{c}_m\hat{c}_m^{\dagger}$) state initially uncoupled to one or more electron reservoirs or leads. The initial condition of the reservoirs is often taken to be their grand canonical distribution.

Accounting for the restrictions on the matrix elements of time-dependent operators discussed in Sec.~\ref{ssec:matrix-elements-of-time-dependent-operators}, when the initial condition is diagonal, the bosonic representation permits the calculation of correlation functions where $\hat{B}(\mathbf{t})$ can assume the forms of the proper-time-ordered operators shown in Eqs.~(\ref{eq:diagonal-matrix-element-general-time-dependent-1-creation-annihilation-pair}) and (\ref{eq:diagonal-matrix-element-general-time-dependent-2-creation-annihilation-pairs}). Examples of these correlation functions include time dependent populations, currents, and their second moments. 

In addition to the dynamical quantities characterized above, there are two special cases which can also be treated using the bosonic representation. First, one can extend the treatment to non-diagonal initial conditions that can be factored into diagonal, $\hat{\rho}_d$, and simple non-diagonal, $\hat{\rho}_{nd}$, components, $\hat{\rho} = \hat{\rho}_{d} \hat{\rho}_{nd}$, that can be included in the time-dependent product, 
\begin{equation}
\begin{split}
    \langle \hat{B}(\{ t_j \}) \rangle &= \sum_{ \{ \mathbf{n} \}}\bra{\mathbf{n}}\hat{\rho}_{d}\ket{\mathbf{n}} \bra{\mathbf{n}}\hat{\rho}_{nd}\hat{B}(\{ t_j \}) \ket{\mathbf{n}}.
\end{split}
\end{equation}
Consistent with the restrictions on the time-dependent operators that a time-dependent operator needs to obey for its matrix elements to be correctly captured in the bosonic representation, the product $\hat{\rho}_{nd}\hat{B}( \mathbf{t})$ needs to be time-proper-ordered at $t=0$. The second case of interest allows one to extend the treatment to cases where the product of $\hat{B}(\mathbf{t})$ consists of a diagonal operator $\hat{B}_j(0)$ and a product of proper-time-ordered operators, $\hat{B}_k(t_1) ... \hat{B}_l(t_N)$. In this case, the diagonal nature of $\hat{B}_j(0)$ allows one to incorporate it into the initial condition, thereby allowing one to probe a product of time-dependent operators that conform to the time-proper-ordered form, i.e., 
\begin{equation}
\begin{split}
    \langle \hat{B}(\mathbf{t}) \rangle &= \sum_{ \{ \mathbf{n} \}}\bra{\mathbf{n}}\hat{\rho}\hat{B}_j(0) \ket{\mathbf{n}} \bra{\mathbf{n}}\hat{B}_k(t_1) ... \hat{B}_l(t_N) \ket{\mathbf{n}}.
\end{split}
\end{equation}
Thus, with the analysis above, we have outlined in what cases the bosonic representation permits the exact calculation of dynamical quantities for both systems in and out of equilibrium. 
\vspace{20pt}

\onecolumngrid

\begin{figure}[t]
    \centering
    \includegraphics[width=0.9\textwidth]{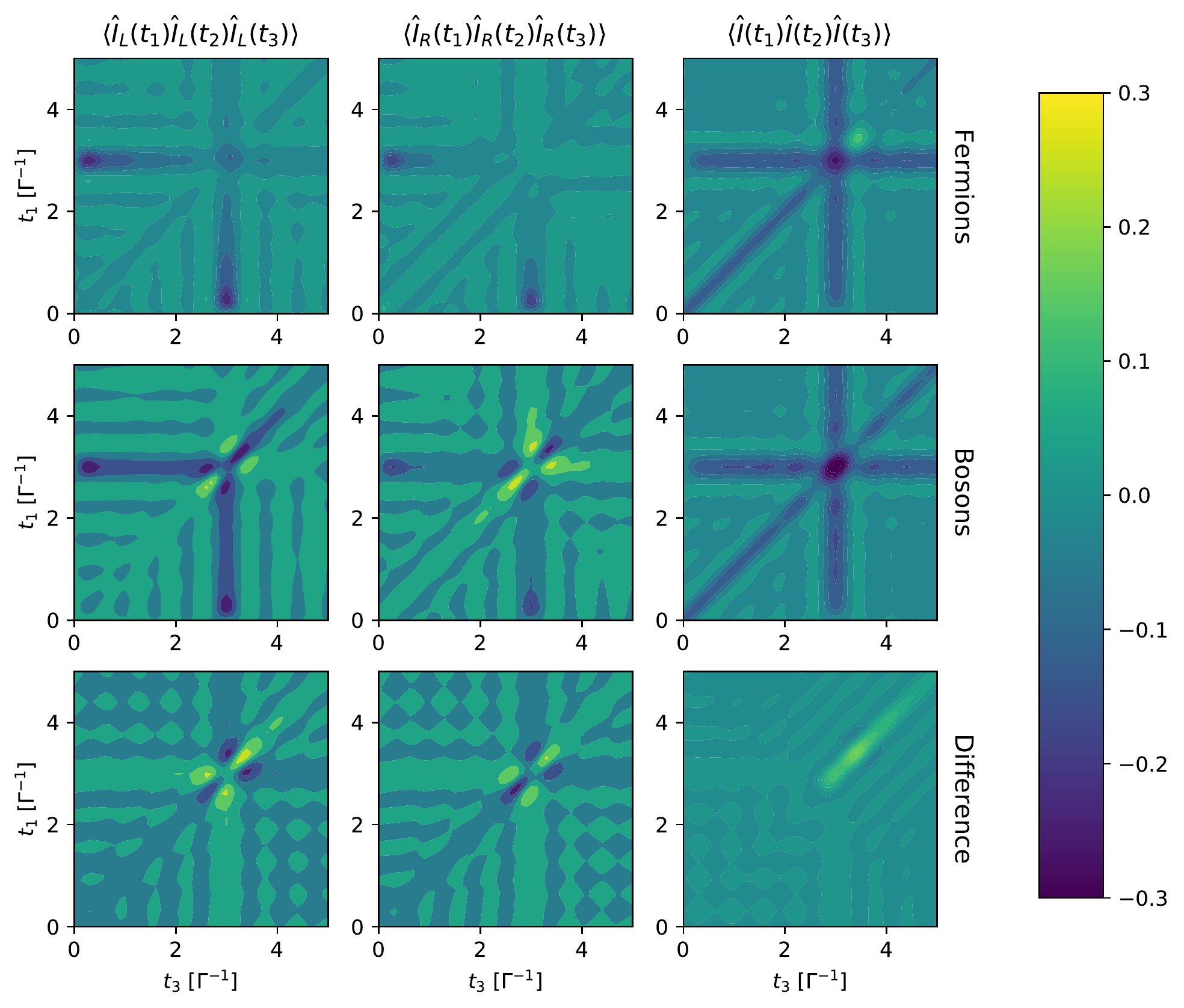}
    \caption{Three-body moments for the resonant level model, where $\varepsilon_{i} = 0$, $\beta_{L} = \beta_{R} = 1/3 \Gamma^{-1}$, $\mu_{L} = -\mu_R = \Gamma$. The second time-index has been held constant at $t_2 = 2.5\Gamma^{-1}$. The leads have been discretized into $N_{L} = N_{R} = 100$ states, and the impurity state is initially assumed to be unoccupied. The top row is the exact result obtained using the fermionic representation, the middle row corresponds to the bosonic result, and the bottom row to the difference between the two: $C_{\rm fer}(t_1, t_2, t_3) - C_{\rm bos}(t_1, t_2, t_3)$ . Each column corresponds to the third moment of different current operators. The leftmost column corresponds to the third moment of the current arising from the left lead, middle column to the current from the right lead, and the rightmost column to the total current. See Eqs.~(\ref{eq:rlm-one-reservoir-current}) and (\ref{eq:rlm-total-current}) for expressions for these currents.} 
    \label{fig:2d-3-body-moments}
\end{figure}

\twocolumngrid

\section{Numerical illustrations}
\label{sec:numerical-illustrations}

To numerically demonstrate the validity of the conclusions derived in Sec.~\ref{sec:quadratic-fermionic-hamiltonian}, here we consider elastic charge transport in a model nanojunction as described via the nonequilibrium resonant level model \cite{MahanBook, HaugJauho}. One of the most widely used forms of the quadratic fermionic Hamiltonian, the resonant level model is conventionally used to model elastic electron transfer from a molecule or quantum dot, commonly called an impurity, to a nearby lead which functions as an electron reservoir, and elastic transport across a nanojunction consisting of an impurity placed between two leads. The Hamiltonian for the resonant level is of the form in Eq.~(\ref{eq:general-quadratic-fermion-hamiltonian}), but can be subdivided into impurity, leads, and impurity-lead coupling components, 
\begin{equation}\label{eq:rlm-hamiltonian}
\begin{split}
    H_{\mathrm{rlm}} &= \sum_{m} \varepsilon_{m} \hat{c}_{m}^{\dagger}\hat{c}_{m} + \sum_{\lambda, k_{\lambda}} \varepsilon_{k_{\lambda}} \hat{c}_{k_{\lambda}}^{\dagger}\hat{c}_{k_{\lambda}} \\
    &\qquad + \sum_{m,\lambda, k_{\lambda}} t_{m,k_{\lambda}} \big( \hat{c}_{m}^{\dagger}\hat{c}_{k_{\lambda}} + \hat{c}_{k_{\lambda}}^{\dagger}\hat{c}_{m} \big),
\end{split}
\end{equation}
where the first term corresponds to the impurity part of the system, the second term to the lead(s), and the third connects the impurity and leads. As such,  $\varepsilon_m$ is the energy of an electron on the $m$th impurity state, $\varepsilon_{k_{\lambda}}$ is the energy of the states in the free electron $\lambda$th reservoir, and $t_{m,k_{\lambda}}$ is the hybridization between the impurity and reservoir states which is responsible for electron hopping from the impurity to the reservoir and vice versa. 

Here we use the resonant level model as a numerical testing ground for the insights developed in later parts of the paper. To fully characterize the model, we take the wide-band limit \cite{HaugJauho} with sharp cutoffs at high and low energy values for the hybridization between the impurity and leads, 
\begin{equation}
\begin{split}
    \Gamma_{m, \lambda}(\varepsilon) &= 2\pi \sum_{k_{\lambda}} |t_{m, k_{\lambda}}|^2\delta(\varepsilon- \varepsilon_{k_{\lambda}})\\
    &= \frac{\Gamma_{m, \lambda}}{\Big(1 + e^{A(\varepsilon - B/2)}\Big)\Big(1 + A^{-A + B/2}\Big)}.
\end{split}
\end{equation}
For all numerical results presented here, we use $\Gamma_{L} = \Gamma_{R} = \frac{1}{2}$, $\Gamma_{m} = \Gamma_{m,L} + \Gamma_{m,R}$, $\Gamma = \max[\Gamma_{m}]$, $A = 5\Gamma$, and $B = 20\Gamma$. Although this describes a continuum of states, we employ a uniform discretization of the leads into $N_{\lambda}$ states with the hybridization parameters given by, 
\begin{equation}
    t_{m, k_{\lambda}}(\varepsilon_{k_{\lambda}}) = \sqrt{\frac{\Gamma_{m,k_{\lambda}} \Delta \varepsilon_{m, k_{\lambda}}}{2\pi} },
\end{equation}
where $\Delta \varepsilon_{m, \lambda} = 2B / (N_{\lambda}-1)$ is the energy difference between adjacent lead states.

While our subsequent discussion is general and applies to a diverse set of observables, we demonstrate the validity of our conclusions with a few observables of interest in systems conventionally modelled using the resonant level model, including the population on impurity states
\begin{equation}\label{eq:rlm-impurity-population}
    P_m(t) = \mathrm{Tr}[\rho \hat{c}_m^{\dagger}(t)\hat{c}_m(t)],
\end{equation}
which are useful in studying the rate of elastic electron transfer at an electrochemical interface. When considering electron transport across a junction connected to two (or more) reservoirs, the current coming into the $\lambda$th lead takes the form, 
\begin{equation}\label{eq:rlm-one-reservoir-current}
\begin{split}
    I_{\lambda}(t) &= \frac{d}{dt}\mathrm{Tr}[\rho \hat{N}_{\lambda}(t)]\\
    &= \mathrm{Tr}[\rho \hat{I}_{\lambda}(t)],
\end{split}
\end{equation}
for $\lambda \in \{\mathrm{R}, \mathrm{L}\}$. In Eq.~(\ref{eq:rlm-one-reservoir-current}), the total occupation and current operators for lead $\lambda$ take the forms,
\begin{subequations}
\begin{align}
    \hat{N}_{\lambda} &= \sum_{k_{\lambda}} \hat{c}_{k_{\lambda}}^{\dagger}\hat{c}_{k_{\lambda}}, \label{eq:}\\
    \hat{I}_{\lambda} &= i[\hat{H}_{\mathrm{rlm}},\hat{N}_{\lambda}] = i \sum_{m, k_{\lambda}} t_{m, k_{\lambda}} \big[ \hat{c}_{m}^{\dagger}\hat{c}_{k_{\lambda}} - \hat{c}_{k_{\lambda}}^{\dagger}\hat{c}_{m}\big].
\end{align}
\end{subequations}
Finally, the total current coming from the left reservoir, going through the junction or impurity, and into the right reservoir takes the form, 
\begin{equation}\label{eq:rlm-total-current}
\begin{split}
    I(t) &= \frac{1}{2}\big( I_R(t) - I_L(t) \big).
\end{split}
\end{equation}
In Eqs.~(\ref{eq:rlm-impurity-population}) and (\ref{eq:rlm-one-reservoir-current}), the trace operation $\mathrm{Tr}[...] = \sum_{\{ \mathbf{n} \}} \bra{\mathbf{n}} ... \ket{\mathbf{n}}$ is done over all $2^M$ many-body states $\{ \ket{\mathbf{n}} \}$ in the Hilbert space, where $\mathbf{n} = (n_1, n_2, ..., n_{M})$ is the vector that specifies the occupation of each single-particle state. 

\begin{figure}[t]
    \centering
    \includegraphics[width=\columnwidth]{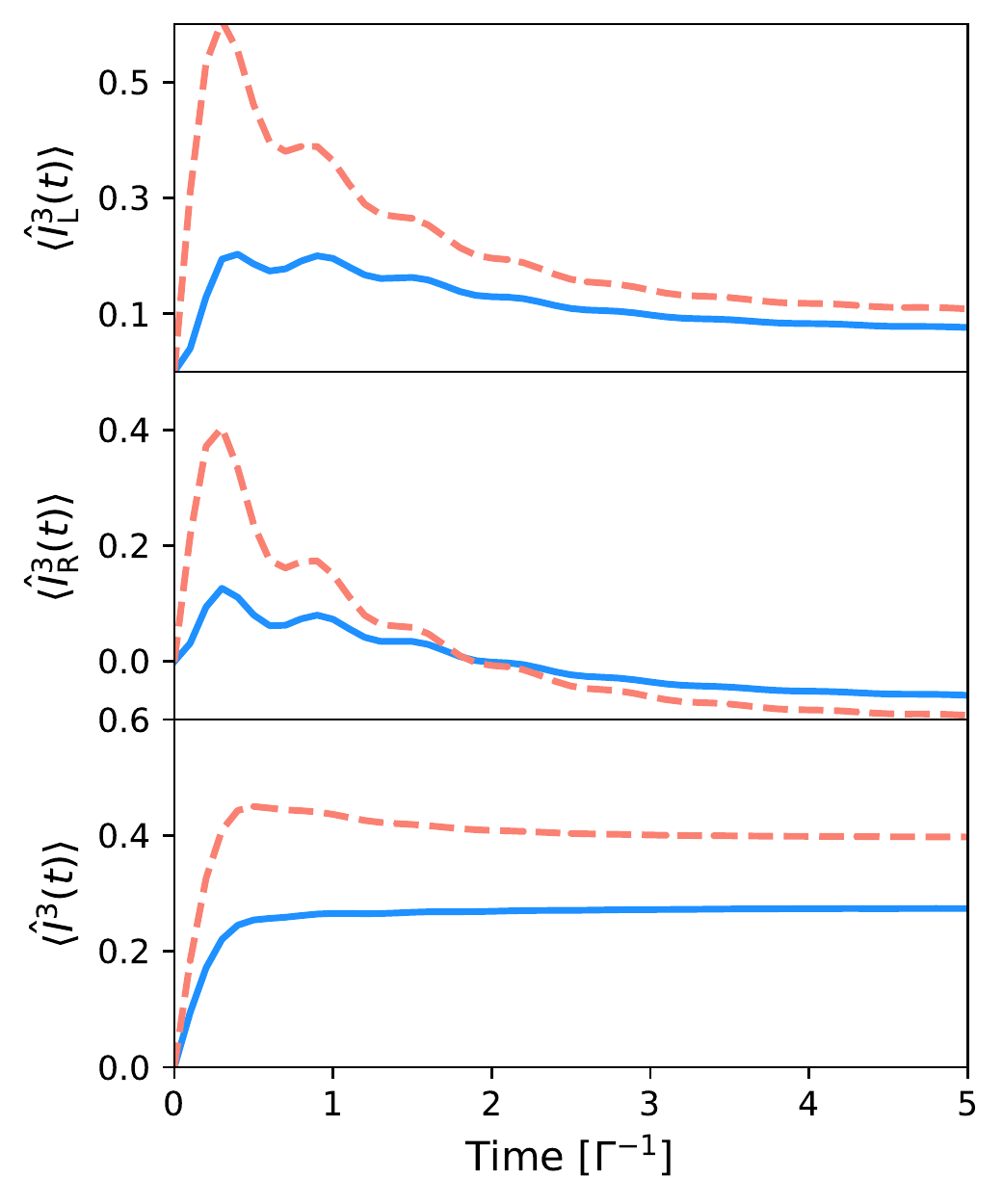}
    \caption{Single-time three-body moments for the same parameterization of the resonant level model in Fig.~\ref{fig:2d-3-body-moments}. The top panel corresponds to the current arising from the left lead; the middle panel to the current from the right lead; and d) the bottom panel to the total current.} 
    \label{fig:rlm-3-body-moments}
\end{figure}

The initial condition for nonequilibrium charge transport in nanojunctions often takes the form, 
\begin{equation}\label{eq:rlm-initial-condition}
    \hat{\rho} = \hat{\rho}_i \prod_{\lambda} \otimes\ \hat{\rho}_{\lambda} .
\end{equation}
The impurity is often assumed to be in its occupied or unoccupied states, $\hat{\rho}_i = \prod_{m} \otimes\ \hat{\rho}_{m}$, where $\hat{\rho}_m = \hat{c}_m^{\dagger}\hat{c}_m$ or $\hat{\rho}_m = \hat{c}_m\hat{c}_m^{\dagger}$, respectively, while the leads in thermal and chemical equilibrium, i.e., $\hat{\rho}_{\lambda} = e^{-\beta(\hat{H}_{\lambda} - \mu_{\lambda}\hat{N}_{\lambda})}/\mathrm{Tr}_{\lambda}[e^{-\beta(\hat{H}_{\lambda} - \mu_{\lambda}\hat{N}_{\lambda})}] = \prod_{k_{\lambda}}\rho_{k_{\lambda}}$.

Figure~\ref{fig:rlm-3-body-moments} shows the third moments of the impurity populations and currents, subject to a diagonal initial condition of the form in Eq.~(\ref{eq:rlm-initial-condition}). Third moments of one-body operators such as those shown in this figure are of a form that require the evaluation of three creation-annihilation pairs of the form given by Eq.~(\ref{eq:fermionic-three-pair-expectation}), which immediately suggests that, even when calculating exact quantum dynamics, the bosonic representation should lead to deviations from the expected fermionic result. While a priori knowledge of this failure does not necessarily suggest that the deviation will be significant, the results in Fig.~\ref{fig:rlm-3-body-moments} demonstrate that the results can be markedly different. Importantly, the third moment of one-body operators, such as those shown in Fig.~\ref{fig:rlm-3-body-moments}, are important dynamical quantities necessary for the calculation of, for example, third-order nonlinear spectroscopies. 

\section{Conclusions}

Here we have shown how direct replacement of fermionic creation and annihilation operators can form an exact quantum mechanical map for systems described by quadratic fermionic Hamiltonians. Specifically, we have shown when this using this map with a restricted Hilbert space corresponding to the physical fermionic Hilbert space can be used to evaluate matrix elements of a wide class of operators. In particular, we have determined the criteria that static and time-dependent operators need to satisfy for the bosonic representation to be able to produce correct matrix elements. 

We have also demonstrated the ability of our analytical framework to determine when one can use the bosonic representation to calculate time-dependent observables, including nonequilibrium and equilibrium single- and multi-time correlation functions. In particular, we have shown that when working with diagonal initial conditions the bosonic representation fails to capture greater than $2$-body time-dependent observables, such as the third moments of the populations and currents in the resonant level model. When initial conditions contain off-diagonal elements, the bosonic representation is able to capture only up to $1$-body time-dependent observables.
    
The analysis presented here establishes the best case scenario limitations for the application of simple bosonic maps to fermionic problems and opens the door for the controlled combination of these maps with the quantum-classical hierarchy. In addition, the analytical framework presented here lays the foundation for the extension of these simple bosonic maps to more complex situations, including those where fermionic degrees of freedom are coupled to complex nuclear motions and may exhibit correlation effects. 

\section*{Acknowledgments}
This work was supported by National Science Foundation Grant No. CHE-2154291. A.M.C.~acknowledges the start-up funds from the University of Colorado, Boulder. 

\section*{Data Availability}
The data that support the findings of this study are available from the corresponding author upon reasonable request.

\appendix

\section{Independent electron approximation}
\label{app:independent-electron-approximation}

A closely related approach used to simplify many-fermion problems described by quadratic fermionic Hamiltonians is the independent electron approximation \cite{Newns1969, Sebastian1989, smithElectronicFrictionElectron1993, Boroda1996a, Thoss2004a}. This approximation replaces creation-annihilation pair products with their single-particle orbitals (see Eq.~(\ref{eq:iea-creation-annihilation-pair-map})), 
\begin{equation}
    \hat{c}_{j}^{\dagger}\hat{c}_k \quad ``\mapsto" \quad  \ket{j}\bra{k}. 
\end{equation}
A direct consequence of this approximation, which transforms the original many-body problem into a one-body multi-state problem, is that the trace operation also changes, 
\begin{equation}\label{eq:iea-trace-map}
    \sum_{ \{\mathbf{n} \} } \bra{\mathbf{n}} ... \ket{\mathbf{n}} \mapsto \sum_{j=1}^{M} \bra{j} ... \ket{j},
\end{equation}
where the vector $\ket{\mathbf{n}}$ is a particular instance of a many-body state in a $2^N$ dimensional Hilbert space, whereas $\ket{j}$ is one single-particle state in the $M$ dimensional Hilbert space for the $M$-level system.

Using Eq.~(\ref{eq:iea-creation-annihilation-pair-map}) the many-fermion in Eq.~(\ref{eq:general-quadratic-fermion-hamiltonian}) becomes the single-particle Hamiltonian, $\mathbf{h}$, which is an $M\times M$ matrix in the basis of the single-particle orbitals, 
\begin{equation}\label{eq:nea-mapped-general-quadratic-fermion-hamiltonian}
\begin{split}
    H \quad ``\mapsto" \quad \mathbf{h} = \sum_{j,k} h_{j,k} \ket{j}\bra{k}.
\end{split}
\end{equation}
As shown in Sec.~\ref{sec:quadratic-fermionic-hamiltonian}, one can diagonalize this single-particle Hamiltonian matrix with the same unitary transformations used for the many-fermion problem to obtain $\mathbf{E} = \mathbf{U}^{-1}\mathbf{h}\mathbf{U}$. This allows one to calculate the time-evolution of an arbitrary one-body fermionic operator
\begin{equation}\label{eq:iea-fermionic-time-dependent-operator}
\begin{split}
    \mathcal{O}^{1}(t) &= \sum_{j,k} \hat{c}_{j}^{\dagger}(t)O_{j,k}\hat{c}_{k}(t)\\
    &= \sum_{r,s} \mathcal{C}_{r,s}^{O}(t) \hat{c}^{\dagger}_{r}\hat{c}_{s}
\end{split}
\end{equation}
as
\begin{equation}\label{eq:iea-mapped-fermionic-time-dependent-operator}
\begin{split}
    \mathcal{O}^{1}(t) &\mapsto \sum_{j,k} O_{j,k} [\ket{j}\bra{k}](t)\\
    &= \sum_{r,s} \mathcal{C}_{r,s}^{O}(t) \ket{r}\bra{s},
\end{split}
\end{equation}
where 
\begin{equation}
    \mathcal{C}_{r,s}^{O}(t) \equiv \sum_{j,k} G_{r,j}^{*}(t)O_{j,k}G_{k,s}(t),
\end{equation}
and $G_{r,s}(t)$ is given by Eq.~(\ref{eq:fermionic-G-time-dependent-matrix}). As Eqs.~(\ref{eq:iea-fermionic-time-dependent-operator}) and (\ref{eq:iea-mapped-fermionic-time-dependent-operator}) demonstrate, the time-dependence of one-body operators can be obtained exactly within the independent electron approximation. 

While the independent electron approximation correctly captures the time-dependence of one-body operators, it requires one to change important details in the way that one calculates observables for many-body problems. To see the relevant differences, we consider the nonequilibrium average of the one-body operator in Eq.~(\ref{eq:iea-fermionic-time-dependent-operator}) in its original second quantized formulation, 
\begin{equation}\label{eq:expectation-value-of-one-body-operator}
\begin{split}
    \langle \mathcal{O}^{1}(t) \rangle &= \sum_{r,s} \mathcal{C}_{r,s}^{O}(t) \mathrm{Tr}[\rho \hat{c}^{\dagger}_{r}\hat{c}_{s} ].
\end{split}
\end{equation}
Using Eqs.~(\ref{eq:iea-trace-map}) and (\ref{eq:nea-mapped-general-quadratic-fermion-hamiltonian}), one obtains the following expression for the mapping of Eq.~(\ref{eq:expectation-value-of-one-body-operator}),
\begin{equation}\label{eq:iea-mapped-expectation-value-of-one-body-operator}
\begin{split}
    \langle \mathcal{O}^{1}(t) \rangle &\mapsto \sum_{r,s} \mathcal{C}_{r,s}^{O}(t) \mathrm{Tr}_{iea}[\tilde{\rho} \ket{r}\bra{s} ],
\end{split}
\end{equation}
where $\tilde{\rho}$ is the mapped density. Clearly, for this mapping to be valid, the many-electron density must be mapped to the sum of one-electron operators. To illustrate this, we consider a common nonequilibrium initial condition for electron conduction in the resonant level model corresponding to the non-interacting bath at thermal equilibrium and an occupied central impurity:
\begin{equation}\label{eq:initial-condition-rlm}
    \rho_{\rm rlm}(0) = \hat{c}^{\dagger}_{0}\hat{c}_{0} \prod_{k} \frac{e^{-\beta (\varepsilon_k - \mu) \hat{c}^{\dagger}_{k}\hat{c}_{k} }}{\mathrm{Tr}_k[e^{-\beta (\varepsilon_k - \mu) \hat{c}^{\dagger}_{k}\hat{c}_{k}}]},
\end{equation}
where the subscript $0$ denotes the impurity and $k \neq 0$ the states in the fermionic bath. Substituting Eq.~(\ref{eq:initial-condition-rlm}) into Eq.~(\ref{eq:expectation-value-of-one-body-operator}), one obtains, 
\begin{equation}\label{eq:expectation-value-of-one-body-operator-substituted}
\begin{split}
    \langle \mathcal{O}^{1}(t) \rangle &= \sum_{r,s} \mathcal{C}_{r,s}^{O}(t) \delta_{r,s} f_{r},
\end{split}
\end{equation}
where 
\begin{equation}
    f_{r} = \begin{cases}
    1 &\ \mathrm{\ for \ } r=0,\\
    \frac{e^{-\beta(\varepsilon_k - \mu)}}{1 + e^{-\beta(\varepsilon_k - \mu)}} &\ \mathrm{\ for \ } r\neq 0.
    \end{cases}
\end{equation}
This example allows us to identify the mapped density, $\tilde{\rho}$, as
\begin{equation}
    \tilde{\rho}_{\rm rlm} = \sum_{r} f_r \ket{r}\bra{r}.
\end{equation}
Interestingly, this density matrix is normalized to the average particle number in the system, $N$, rather than $1$. Since the density must be mapped to a one-body operator in the independent electron approximation, the many-body problem should be rotated to a basis where the full density can be written as the product of one-electron operators that can be mapped using a similar procedure as in the resonant-level model above. 

Attempts to go beyond one-body operators illustrate some of the limitations of the independent electron approximation. We consider a generic two-body operator as an example, 
\begin{equation}\label{eq:iea-fermionic-time-dependent-2-body-operator}
\begin{split}
    \mathcal{O}^{2} &= \sum_{j,k,l,m} O_{j,k,l,m}\hat{c}^{\dagger}_{j}\hat{c}^{\dagger}_{k}\hat{c}_{m}\hat{c}_{l}.
\end{split}
\end{equation}
For simplicity, we consider the time-independent case as it illustrates the difficulty in using the independent electron approximation and its time dependence can be obtained trivially using the protocol outlined above for the 1-body operator. Evaluating the average of this operator using the resonant level model example above yields,
\begin{equation}\label{eq:expectation-value-of-two-body-operator-substituted}
\begin{split}
    \langle \mathcal{O}^{2}(t) \rangle &= \sum_{j,k,l,m} O^{(2)}_{j,k,l,m} [\delta_{j,l}\delta_{k,m} - \delta_{j,m}\delta_{k,l} ]f_{j}f_{k}.
\end{split}
\end{equation}
Yet, this result is incompatible with the mapping in Eq.~(\ref{eq:iea-creation-annihilation-pair-map}), which would require decomposing the two-body operator into one-body pairs that could be mapped unambiguously, say
\begin{equation}
\begin{split}
    \hat{c}^{\dagger}_{j}\hat{c}^{\dagger}_{k}\hat{c}_{m}\hat{c}_{l} &= -(\hat{c}^{\dagger}_{j}\hat{c}_{m})(\hat{c}^{\dagger}_{k}\hat{c}_{l})\\
    &``\mapsto'' \ -\ket{j}\braket{m|k}\bra{l}\\
    &=  -\ket{j}\bra{l}\delta_{m,k},
\end{split}
\end{equation}
or
\begin{equation}
\begin{split}
    \hat{c}^{\dagger}_{j}\hat{c}^{\dagger}_{k}\hat{c}_{m}\hat{c}_{l} &= (\hat{c}^{\dagger}_{j}\hat{c}_{l})(\hat{c}^{\dagger}_{k}\hat{c}_{m})\\
    &``\mapsto'' \ -\ket{j}\braket{l|k}\bra{m}\\
    &=  -\ket{j}\bra{m}\delta_{l,k}.
\end{split}
\end{equation}
Thus, while the independent electron approximation can capture the correct expectation value of one-body operators subject to initial conditions constructed from the product of one-body densities, it cannot be used to calculate the expectation value of many-body operators or the multi-time correlation functions of one-body operators. 

\section{Proof: pairwise ordering for static operators}
\label{sec:proof-pairwise-ordered-operators-static-matrix-elements}

Here, we prove that when a static operator consists of products of single-mode creation-annihilation pairs that appear only once, the property that determines whether one can calculate its matrix element in the bosonic representation is if the operator can be made pairwise ordered in an even number of permutations. 

To prove this statement, we first show that action of a pairwise ordered operator never leads to negative (occupation-dependent) phases. To do this, it is sufficient to consider the action of a one single-mode creation-annihilation pair acting on an arbitrary many-body state, $\ket{\mathbf{n}}$,
\begin{subequations}\label{eq:action-of-single-creation-annihilation-pair-on-many-body-state}
\begin{align}
    \hat{c}^{\dagger}_j\hat{c}_j\ket{\mathbf{n}} &= (-1)^{2h(j)}\delta_{n_j,1}\ket{\mathbf{n}} = \delta_{n_j,1}\ket{\mathbf{n}},\\
    \hat{c}_j\hat{c}^{\dagger}_j\ket{\mathbf{n}} &= (-1)^{2h(j)}\delta_{n_j,0}\ket{\mathbf{n}} = \delta_{n_j,0}\ket{\mathbf{n}}.
\end{align}
\end{subequations}
Note that single-mode creation-annihilation pairs are just occupation operators, $\hat{n}_j = \hat{c}^{\dagger}_j\hat{c}_j\ket{\mathbf{n}} = 1 - \hat{c}_j\hat{c}^{\dagger}_j\ket{\mathbf{n}}$, and the many-body states in the occupation basis form their eigenbasis. As such, the action of creation-annihilation pairs does not modify the many-body state itself, other than acquiring a weight of zero or one depending on the specific occupation of the $j$th mode. Besides being diagonal, these operators lead to no negative phases due to their immediate proximity, i.e., there are no additional creation or annihilation operators corresponding to other modes in between those forming the pair of interest that could change occupation of other single-particle modes and hence prevent the doubling of each operator's contribution to the phase in Eq.~(\ref{eq:action-of-single-creation-annihilation-pair-on-many-body-state}). Because for the purpose of this proof we are only interested in whether the action of creation-annihilation pairs lead to negative phases and, as Eq.~(\ref{eq:action-of-single-creation-annihilation-pair-on-many-body-state}) shows, the order of creation vs annihilation operator in the pair only changes the acquired zero or unity weight $\delta_{n_j,1}$ or $\delta_{n_j, 0}$ corresponding to the occupation of mode $j$, we ignore the identity of individual operators as creation or annihilation operators as long as we consider pairs corresponding to a single mode. Thus, we represent both $\hat{c}_j^{\dagger}\hat{c}_j$ and $\hat{c}_j \hat{c}_{j}^{\dagger}$ as $\hat{j}\hat{j}$. Using this notation, it is clear that when we consider the action of an arbitrary product of pairwise ordered operators on a many-body state $\ket{\mathbf{n}}$, one obtains the same state multiplied by a coefficient, $|\alpha| \geq 0$, which does not change the sign of the state,  
\begin{equation}
\begin{split}
    \hat{j}\hat{j}\hat{k}\hat{k}...\hat{l}\hat{l} \ket{\mathbf{n}} &= |\alpha|\ket{\mathbf{n}}.
\end{split}
\end{equation}
Thus, a pairwise ordered operator consisting of a product of single-mode creation-annihilation pairs always leads to diagonal matrix elements with either positive or zero weights. 

The second part of this proof then requires one to show that any operator consisting of products of single-mode creation-annihilation pairs also lead to diagonal elements with positive or zero weights only if these can be rearranged into a pairwise ordered form using an even number of permutations. For such operators, the permutations necessary to achieve pairwise ordering occur only over operators of different indices. For example, to rearrange $\hat{k}\hat{l}\hat{j}\hat{k}\hat{l}\hat{j}$ into a pairwise ordered form, one needs to permute the $\hat{j}$ operator in the middle twice to the right over the $\hat{k}$ and $\hat{l}$ operators, and then the leftmost $\hat{l}$ operator once to the right over its neighboring $\hat{k}$ operator, thus requiring a total of $P=3$ permutations. In addition, the anticommutivity of fermions dictates that every time a permutation of two operators with different indices occurs, the reordered operator acquires a phase of $-1$. Thus,
\begin{equation}\label{eq:permutations-to-pairwise-order}
\begin{split}
    \hat{k}\hat{l}\hat{j}\hat{k}\hat{l}\hat{j} &= (-1)^{P=2} \hat{k}\hat{l}\hat{k}\hat{l}\hat{j}\hat{j}\\
    &= (-1)^{P=3} \hat{k}\hat{k}\hat{l}\hat{l}\hat{j}\hat{j},
\end{split}
\end{equation}
where the exponent of the phase, $-1$, accounts for the number of permutations necessary to bring the operator into a pairwise ordered form. In its pairwise ordered form, evaluation of the matrix elements of the operator acquire no additional negative phases. Therefore, the matrix element of the operator type in Eq.~(\ref{eq:permutations-to-pairwise-order}) contains a negative phase which cannot be captured by the bosonic representation,
\begin{equation}
\begin{split}
    \bra{\mathbf{n}} \hat{k}\hat{l}\hat{j}\hat{k}\hat{l}\hat{j} \ket{\mathbf{n}} &= (-1)^{3}
\end{split}
\end{equation}
Here, we have chosen a given configuration of the creation and annihilation operators corresponding to the generic order given by Eq.~(\ref{eq:permutations-to-pairwise-order}). In contrast, the bosonic representation is able to capture the matrix elements of operators which only require an even number of permutations, such as, 
\begin{equation}\begin{split}
    \hat{l}\hat{k}\hat{j}\hat{k}\hat{l}\hat{j} &= (-1)^{P=2} \hat{l}\hat{k}\hat{k}\hat{l}\hat{j}\hat{j}\\
    &= (-1)^{P=4} \hat{k}\hat{k}\hat{l}\hat{l}\hat{j}\hat{j}.
\end{split}
\end{equation}
which do not acquire negative phases. 

A simple yet important consequence of the fact that no negative phases emerge when evaluating the diagonal matrix elements proper-ordered operators, i.e., operators consisting of products of single-mode creation-annihilation pairs in any order that leads to pairwise ordering in an even number of permutations, is that powers of proper-ordered operators are also proper-ordered. To see this, it is sufficient to consider the action of the operator $\mathcal{O}$ on a many-body state in the occupation basis, 
\begin{equation}\label{eq:action-of-proper-ordered-operator-on-ket}
    \mathcal{O}\ket{\mathbf{n}} = |\alpha_{\mathcal{O}}|\ket{\mathbf{n}},
\end{equation}
where, as was shown above, results in a weight that is zero or positive, $|\alpha_{\mathcal{O}}| \geq 0$ and which encodes the occupation of the modes appearing in the operator $\mathcal{O}$. Using Eq,.~(\ref{eq:action-of-proper-ordered-operator-on-ket}), one can see that the action of any power of $\mathcal{O}$ on the same many-body state also results in a weight that is positive or zero,  
\begin{equation}
\begin{split}
    \mathcal{O}^m \ket{\mathbf{n}} &= |\alpha_{\mathcal{O}}|^{m}\ket{\mathbf{n}}.
\end{split}
\end{equation}

Hence, \textit{when an operator consisting of products of single-mode creation-annihilation pairs arranged in an arbitrary order requires an even number of permutations to bring it to a pairwise ordered form, its diagonal matrix elements do not acquire negative phases. These operators and their powers, denoted as proper-ordered for compactness and clarity, are therefore compatible with the bosonic representation.} Conversely, when an operator of the type described above requires an odd number of permutations to achieve a pairwise ordered form, evaluation of its matrix elements result in a negative phase, which the bosonic representation does not capture.

\section{Relation to previous quasiclassical maps}
\label{app:other-maps}

In this Appendix, we show that several previous quasiclassical maps for many-fermion problems \cite{Li2012, Li2014b, Levy2019} follow bosonic statistics and that the conclusions we draw in Sec.~\ref{sec:quadratic-fermionic-hamiltonian} are applicable, allowing us to understand on a rigorous footing the source of their success and outline their limitations.

\subsubsection{Li-Miller-Levy-Rabani map}

We begin with the quasiclassical map presented in Ref.~\onlinecite{Li2014b}, where fermionic operators are mapped as
\begin{subequations}
\begin{align}
    \hat{c}_{j}^{\dagger} \ \ ``&\mapsto" \ \ \frac{1}{2}\Big((\hat{q}_{xj} + \hat{p}_{yj}) +i (\hat{q}_{yj} - \hat{p}_{xj})\Big) \nonumber \\
    &\qquad \qquad = (\hat{b}^{\dagger}_{xj} + i \hat{b}_{yj}^{\dagger})/\sqrt{2}\\
    \hat{c}_{j} \ \ ``&\mapsto"\ \ \frac{1}{2}\Big((\hat{q}_{xj} + \hat{p}_{yj}) - i (\hat{q}_{yj} + \hat{p}_{xj})\Big) \nonumber \\
    &\qquad \qquad = (\hat{b}_{xj} - i \hat{b}_{yj})/\sqrt{2}.
\end{align}
\end{subequations}
Although it may initially appear that this mapping is distinct from that explored in this work, one can define new bosonic operators
\begin{subequations}
\begin{align}
    \hat{b}_{j} &\equiv \frac{1}{\sqrt{2}}(\hat{b}_{xj} - i \hat{b}_{yj}),\\
    \hat{b}_{j} &\equiv \frac{1}{\sqrt{2}}(\hat{b}_{xj} - i \hat{b}_{yj}),
\end{align}
\end{subequations}
which obey the bosonic commutation relations in Eq.~(\ref{eq:bosonic-commutation-relation}). Thus, one can simplify the mapping in Ref.~\onlinecite{Li2014b} to 
\begin{subequations}
\begin{align}
    \hat{c}_{j}^{\dagger} \ \ ``&\mapsto" \ \  \hat{b}^{\dagger}_{j},\\
    \hat{c}_{j} \ \ ``&\mapsto"\ \ \hat{b}_{j},
\end{align}
\end{subequations}
which is equivalent to the map in Eq.~(\ref{eq:fermion-to-boson-inexact-map}). This means that the analysis provided in the main part of the paper applies directly to this map. 

\subsubsection{Li-Miller \& Levy-Dou-Rabani-Limmer maps}

We can perform a similar analysis of the mapping approach presented in Ref.~\onlinecite{Li2012} and revisited recently in Ref.~\onlinecite{Levy2019}. In these quasiclassical mappings, quadratic products of creation and annihilation operators are mapped as, 
\begin{subequations}
\begin{align}
    \hat{c}_n^{\dagger}\hat{c}_m &\mapsto \frac{1}{2}\Big[ q_{xn}p_{ym} - p_{xn}q_{ym} + p_{yn}q_{xm} - q_{yn}p_{xm} \nonumber \\ 
    &\qquad  + i \Big( q_{xn}p_{xm} - p_{xn}q_{xm} + q_{yn}p_{ym} - p_{yn}q_{ym} \Big) \Big].
\end{align}
\end{subequations}
Quantizing these classical variables, we can translate the Cartesian operators into combinations of bosonic operators,
\begin{equation}\label{eq:li-miller-map}
\begin{split}
    \hat{c}_n^{\dagger}\hat{c}_m &\mapsto \frac{1}{2}\Big[ \hat{q}_{xn}\hat{p}_{ym} - \hat{p}_{xn}\hat{q}_{ym} + \hat{p}_{yn}\hat{q}_{xm} - \hat{q}_{yn}\hat{p}_{xm} \\ 
    &\qquad  + i \Big( \hat{q}_{xn}\hat{p}_{xm} - \hat{p}_{xn}\hat{q}_{xm} + \hat{q}_{yn}\hat{p}_{ym} - \hat{p}_{yn}\hat{q}_{ym} \Big) \Big] \\
    &= \frac{1}{2}\Big[ (\hat{b}_{yn}^{\dagger} - i\hat{b}_{xn}^{\dagger})(\hat{b}_{ym} + i \hat{b}_{xm}) \\
    &\qquad \quad -(\hat{b}_{yn} - i\hat{b}_{xn})(\hat{b}_{ym}^{\dagger} + i \hat{b}_{xm}^{\dagger}) \Big] \\
    &\equiv  \hat{b}_{\alpha n}^{\dagger}\hat{b}_{\alpha m} - \hat{b}_{\beta n}\hat{b}_{\beta m}^{\dagger},
\end{split}
\end{equation}
where
\begin{subequations}
\begin{align}
    \hat{b}_{\alpha n} &= \frac{1}{\sqrt{2}}(\hat{b}_{yn} + i \hat{b}_{xn}),\\
    \hat{b}_{\alpha n}^{\dagger} &= \frac{1}{\sqrt{2}}(\hat{b}^{\dagger}_{yn} - i \hat{b}^{\dagger}_{xn}),\\
    \hat{b}_{\beta n} &= \frac{1}{\sqrt{2}}(\hat{b}_{yn} - i \hat{b}_{xn}),\\
    \hat{b}_{\beta n}^{\dagger} &= \frac{1}{\sqrt{2}}(\hat{b}^{\dagger}_{yn} + i \hat{b}^{\dagger}_{xn}).
\end{align}
\end{subequations}
Since the $\alpha$ and $\beta$ modes arise from orthogonal combinations of the $x$ and $y$ modes, these are orthogonal, independent bosons, which means that they follow conventional commutation relations, 
\begin{subequations}
\begin{align}
    [\hat{b}_{\gamma n}, \hat{b}^{\dagger}_{\gamma \prime m}] &= \delta_{\gamma, \gamma \prime}\delta_{n,m},\\
    [\hat{b}_{\gamma n}, \hat{b}_{\gamma \prime m}] &= 0 = [\hat{b}^{\dagger}_{\gamma n}, \hat{b}_{\gamma \prime m}].
\end{align}
\end{subequations}

Although in this quasiclassical turned quantum mechanical map individual fermionic creation and annihilation operators are not directly mapped to bosonic ones, we demonstrate that the major conclusions that we draw in Sec.~\ref{sec:quadratic-fermionic-hamiltonian} also apply to this map. To do this, we first note that, writing a quadratic Hamiltonian of the form in Eq.~(\ref{eq:general-quadratic-fermion-hamiltonian}) in terms of the map given by Eq.~(\ref{eq:li-miller-map}), one can separate the Hamiltonian into two commuting terms corresponding to the $\alpha$ and $\beta$ modes,
\begin{equation}
\begin{split}
    H \mapsto \mathcal{H}^{\alpha} + \mathcal{H}^{\beta},
\end{split}
\end{equation}
where
\begin{subequations}
\begin{align}
    \mathcal{H}^{\alpha} &= \sum_{j,k} h_{j,k} \hat{b}_{j \alpha}^{\dagger}\hat{b}_{k \alpha},\\
    \mathcal{H}^{\beta} &= - \sum_{j,k} h_{j,k} \hat{b}_{j \beta}\hat{b}_{k \beta}^{\dagger}.
\end{align}
\end{subequations}

To obtain the time-dependence of arbitrary operators, we first calculate the time-dependence of individual bosonic creation and annihilation operators. Using the same transformations as those used in Sec.~\ref{sec:quadratic-fermionic-hamiltonian}, one can show that
\begin{subequations}\label{eq:li-miller-time-dependence-bosons}
\begin{align}
    \hat{b}^{\dagger}_{\alpha j}(t) = \sum_{k} G_{k,j}^{*}(t) \hat{b}^{\dagger}_{\alpha k},\\
    \hat{b}_{\alpha j}(t) = \sum_{k} G_{j,k}(t) \hat{b}_{\alpha k},\\
    \hat{b}^{\dagger}_{\beta j}(t) = \sum_{k} G_{j,k}(t) \hat{b}^{\dagger}_{\beta k},\\
    \hat{b}_{\beta j}(t) = \sum_{k} G_{k,j}^{*}(t) \hat{b}_{\beta k},
\end{align}
\end{subequations}
where $G_{j,k}(t)$ is given by Eq.~(\ref{eq:fermionic-G-time-dependent-matrix}). 

Equation~(\ref{eq:li-miller-time-dependence-bosons}) allows one to construct the time-dependence of any operator. However, since the original quasiclassical map of Ref.~\onlinecite{Li2012} provides expressions only for quadratic operators, we begin our analysis with such operators. When using the Li-Miller map in the quantum mechanical form provided in the last line of Eq.~(\ref{eq:li-miller-map}), one can separate time-dependent one-body operators into contributions arising from the $\alpha$ and $\beta$ modes, 
\begin{equation} \label{eq:li-miller-one-body-operator}
\begin{split}
    \mathcal{O}^{1}(t) &= \sum_{j,k} \hat{c}_{j}(t)O_{j,k}\hat{c}_{k}(t)\\
    &= \mathcal{O}^{1\alpha}(t) + \mathcal{O}^{1\beta}(t),
\end{split}
\end{equation}
where
\begin{subequations}
\begin{align}
    \mathcal{O}^{1\alpha}(t) &= \sum_{j,k} \hat{b}^{\dagger}_{j\alpha}(t)O_{j,k}\hat{b}_{k\alpha}(t) =  \sum_{r,s} \mathcal{C}^{O}_{r,s}(t) \hat{b}^{\dagger}_{r\alpha}  \hat{b}_{s\alpha} ,\\
    \mathcal{O}^{1\beta}(t) &= -\sum_{j,k} \hat{b}_{j\beta}(t)O_{j,k}\hat{b}^{\dagger}_{k\beta}(t) = -\sum_{r,s} \mathcal{C}^{O}_{r,s}(t) \hat{b}_{r\beta}\hat{b}^{\dagger}_{\beta s}.
\end{align}
\end{subequations}
and
\begin{equation}
    C^{O}_{r,s}(t) \equiv \sum_{j,k} G_{r,j}^{*}(t) O_{j,k} G_{k,s}(t).
\end{equation}
One can therefore express the time-evolved version of this one-body operator as
\begin{equation}
\begin{split}
    \mathcal{O}^{1}(t) &= \sum_{r,s} \Bigg[\sum_{j,k}  G^{*}_{r,j}(t) O_{j,k} G_{k,s}(t)\Bigg] \hat{c}_{r}^{\dagger}\hat{c}_{s}\\
    &= \mathcal{O}^{1\alpha} + \mathcal{O}^{1\beta},
\end{split}
\end{equation}
which demonstrates that this quasiclassical turned quantum mechanical map captures the correct quantum dynamics of many-fermion problems described by quadratic Hamiltonians of the form in Eq.~(\ref{eq:general-quadratic-fermion-hamiltonian}) when applied to quadratic operators. In fact, since the time-dependence of a single creation-annihilation product is captured correctly in this quantum mechanical map, it also correctly captures the time-dependence of any operator consisting of products of single-creation-annihilation pairs. To illustrate this point, we consider the product of two one-body operators of the form in Eq.~(\ref{eq:li-miller-one-body-operator}), 
\begin{equation}\label{eq:li-miller-quartic-operator-evolved-fermion}
\begin{split}
    \mathcal{O}^{2}(t) &= \mathcal{O}^{1}(t)\mathcal{Q}^{1}(t)\\
    &= \sum_{j,k,l,m} O_{j,k}Q_{l,m} \hat{c}_{j}^{\dagger}(t)\hat{c}_{k}(t)\hat{c}_{l}^{\dagger}(t)\hat{c}_{m}(t) \\ 
    &= \sum_{r,s,u,v} \mathcal{C}^{O}_{r,s}(t)\mathcal{C}^{Q}_{u,v}(t) \hat{c}_{r}^{\dagger}\hat{c}_{s}\hat{c}_{u}^{\dagger}\hat{c}_{v},
\end{split}
\end{equation}
where $O$ and $Q$ are arbitrary one-body operators. Upon mapping, the time-dependence of these operators takes the form, 
\begin{equation}
\begin{split}
    \mathcal{O}^{2}(t) &\mapsto [\mathcal{O}^{1 \alpha}(t)+\mathcal{O}^{1 \beta}(t)][\mathcal{Q}^{1\alpha }(t) + \mathcal{Q}^{1\beta}(t)]\\
    &= \sum_{r,s,u,v} \mathcal{C}^{O}_{r,s}(t)\mathcal{C}^{Q}_{u,v}(t) \\
    &\qquad \quad \times \big[\hat{b}_{\alpha r}^{\dagger}\hat{b}_{\alpha s} - \hat{b}_{\beta r}^{\dagger}\hat{b}_{\beta s}\big]\big[\hat{b}_{\alpha u}^{\dagger}\hat{b}_{\alpha v} - \hat{b}_{\beta u}^{\dagger}\hat{b}_{\beta v}\big],
\end{split}
\end{equation}
which exactly corresponds to the result one would obtain by mapping the two creation-annhilation pairs in the last line of Eq.~(\ref{eq:li-miller-quartic-operator-evolved-fermion}). 

To complete our analysis of this map, we provide the physical basis on which it acts, which we derive by considering the action of the fermionic occupation number operator on the occupied and unoccupied states, and of a fermion transfer term on an appropriate state. We begin by applying the fermionic number operator on an occupied state, which should return the same state multiplied by unity,
\begin{equation}
\begin{split}
    \hat{c}_n^{\dagger}\hat{c}_n \ket{1_n}  &= 1\ket{1_n}\\
    &\mapsto (\hat{b}_{\alpha n}^{\dagger}\hat{b}_{\alpha n}  - \hat{b}_{\beta n}^{\dagger}\hat{b}_{\beta n}  - 1)\ket{x_{\beta n}, y_{\alpha n}}\\
    &\mapsto (1 +  x_{\beta n}  - y_{\alpha n})\ket{x_{\beta n}, y_{\alpha n}},
\end{split}
\end{equation}
where $x, y \in \mathbb{Z}$. This implies that
\begin{equation}
    x_{\beta n} = y_{\alpha n}
\end{equation}
for an occupied state. In contrast, upon applying the fermionic occupation number operator to an empty state, one should recover same state multiplied by zero,
\begin{equation}
\begin{split}
    \hat{c}_n^{\dagger}\hat{c}_n \ket{0_n}  &= 0\ket{0_n}\\
    &\mapsto (\hat{b}_{\alpha n}^{\dagger}\hat{b}_{\alpha n}  - \hat{b}_{\alpha n}^{\dagger}\hat{b}_{\beta n}  - 1)\ket{x_{\beta n}, y_{\alpha n}}\\
    &\mapsto (1 +  x_{\beta n}  - y_{\alpha n})\ket{x_{\beta n}, y_{\alpha n}},
\end{split}
\end{equation}
which implies that 
\begin{equation}
    y_{\alpha n} = x_{\beta n} + 1.
\end{equation}
Hence,
\begin{subequations}
\begin{align}
    \ket{1_n} &\mapsto \ket{x_{\beta n}, x_{\alpha n}},\\
    \ket{0_n} &\mapsto \ket{(y-1)_{\beta n}, y_{\alpha n}}.
\end{align}
\end{subequations}
    
While this analysis does not establish the relation between $x$ and $y$, it is straightforward to determine it by considering the action of the quadratic product that removes an electron from the $m$th orbital and places one in the $n$th orbital on the fermionic basis,
\vspace{20pt}
\begin{widetext}
\begin{equation}
\begin{split}
    \hat{c}_n^{\dagger}&\hat{c}_m\ket{1_m, 0_n} = \ket{0_m, 1_n}\\
    &\mapsto (\hat{b}_{\alpha n}^{\dagger}\hat{b}_{\alpha m}  - \hat{b}_{\beta m}^{\dagger}\hat{b}_{\beta n})  \ket{(y-1)_{\beta m}, y_{\alpha m}; x_{\beta n}, x_{\alpha n}}\\
    &=\hat{b}_{\alpha n}^{\dagger}\hat{b}_{\alpha m}\ket{(y-1)_{\beta m}, y_{\alpha m}; x_{\beta n}, x_{\alpha n}}  - \hat{b}_{\beta m}^{\dagger}\hat{b}_{\beta n}  \ket{(y-1)_{\beta m}, y_{\alpha m}; x_{\beta n}, x_{\alpha n}}\\
    &=\sqrt{x+1}\sqrt{y}\ket{(y-1)_{\beta m}, (y-1)_{\alpha m}; x_{\beta n}, (x+1)_{\alpha n}}  -  \sqrt{y}\sqrt{x} \ket{y_{\beta m}, y_{\alpha m}; (x-1)_{\beta n}, x_{\alpha n}}
\end{split}
\end{equation}
\end{widetext}
This expression allows us to determine that $x = 0$ and $y=1$, implying that the physical basis for the Li-Miller map takes the form,
\begin{subequations}
\begin{align}
    \ket{1_n} \mapsto \ket{0_{\beta n}, 0_{\alpha n}},\\
    \ket{0_n} \mapsto \ket{0_{\beta n}, 1_{\alpha n}},
\end{align}
\end{subequations}
where an occupied fermionic orbital is encoded by a fully unoccupied many-body state corresponding to the $\alpha$ and $\beta$ bosons, while the unoccupied fermionic orbital corresponds to the many-body state where the $\alpha$ boson has one excitation and the $\beta$ boson is in the ground state.

\section{Phase space formulation and the classical limit}
\label{sec:phase-space-classical-limit}

Below we demonstrate that for the types of continuous variable Hamiltonians that one obtains from the bosonic representation in Eq.~(\ref{eq:fermion-to-boson-inexact-map}), classical dynamics are certain to recover the exact quantum mechanical result when calculating one-time correlation functions and nonequilibrium averages. 

When the bosonic representation for quadratic fermionic Hamiltonians of the form in Eq.~(\ref{eq:general-quadratic-fermion-hamiltonian}) is also quadratic, as is the case when using Eq.~(\ref{eq:fermion-to-boson-inexact-map}), the resulting bosonic Hamiltonian in Eq.~(\ref{eq:general-quadratic-fermion-hamiltonian-in-bosonic-operators}) can be rewritten in terms of Cartesian operators (see Eq.~(\ref{eq:general-quadratic-fermion-hamiltonian-in-cartesian-operators}), reproduced here),
\begin{equation}
\begin{split}
    \mathcal{H} &= \frac{1}{2} \sum_{j, k} h_{j,k} \Big[\hat{q}_j\hat{q}_k + \hat{p}_j\hat{p}_k - \delta_{j,k} + i \big(\hat{q}_j\hat{p}_k - \hat{p}_j\hat{q}_k \big) \Big],
\end{split}
\end{equation}
where we have used the fact that $\hat{b}_j = (\hat{q}_j + i\hat{p}_j)/\sqrt{2}$ and $\hat{b}^{\dagger}_j = (\hat{q}_j - i\hat{p}_j)/\sqrt{2}$. Although several phase space formulations, corresponding to different operator orderings, exist \cite{Hillery1984a, Polkovnikov2010}, in the following we focus on the Wigner phase space formulation \cite{Hillery1984a}.

In the Wigner phase space formulation, a single-time correlation function (and nonequilibrium averages, when $\rho \neq \rho_{eq}$ and $\hat{A} = \hat{\mathbf{1}}$) takes a particularly straightforward form, 
\begin{equation}\label{eq:correlation-function-in-wigner-phase-space}
\begin{split}
    C_{AB}(t) &= \mathrm{Tr}[\rho \hat{A}(0) \hat{B}(t)]\\
    &= \frac{1}{(2\pi\hbar)^{N}} \int d\mathbf{x} d\mathbf{p}\ [\hat{\rho}\hat{A}]^W(\mathbf{x}, \mathbf{p})[\hat{B}(t)]^W(\mathbf{x}, \mathbf{p}),
\end{split}
\end{equation}
where $N$ is the number of modes which are transformed, the superscript $W$ denotes the Wigner transform of an operator, which takes the form,
\begin{equation}\label{eq:wigner-transform-defintion}
    \hat{A}^W = \int d\mathbf{s} e^{-i\mathbf{p} \cdot \mathbf{s} / \hbar} \bra{\mathbf{x} + \mathbf{s}/2} \hat{A} \ket{\mathbf{x} - \mathbf{s}/2}.
\end{equation}
and the Wigner transform of products of operators
\begin{equation}\label{eq:wigner-transform-of-two-operators}
    (\hat{A}\hat{B})^W = A^W e^{\hbar \hat{\Lambda} / 2i}B^W
\end{equation}
involves the Moyal bracket
\begin{equation}\label{eq:moyal-bracket-definition}
    \hat{\Lambda} = \boldsymbol{\nabla}_{p} \cdot \boldsymbol{\nabla}_{q} - \boldsymbol{\nabla}_{q} \cdot \boldsymbol{\nabla}_{p},
\end{equation}
which is of the same form as the classical Poisson bracket, i.e., $A^W \hat{\Lambda} B^W = - \{A^W, B^W \}_{PB}$.

To calculate the time-dependence of a Wigner transformed operator, $[\hat{B}(t)]^W$, one can Wigner transform its quantum mechanical equation of motion, 
\begin{equation}
\begin{split}
    \frac{d}{dt}[\hat{B}(t)]^W &= i\Big[ [\hat{H}, \hat{B}(t)] \Big]^W\\
    &= [\hat{H}]^W \Bigg[\hat{\Lambda} + \frac{(\hbar/i)^2}{3!}\hat{\Lambda}^3 + ... \Bigg][\hat{B}(t)]^W.
\end{split}
\end{equation}
When the Hamiltonian, $H^W(\mathbf{x}, \mathbf{p})$, is at most quadratic in the positions and momenta, only the first term in the expansion above yields a finite contribution, i.e., $H^W\hat{\Lambda}^{n}[\hat{B}(t)]^W = 0$ for $n \geq 3$. This  renders the quantum mechanical equation of motion of an arbitrary Wigner-transformed operator $[\hat{B}(t)]^W$ equivalent to its classical equation of motion, 
\begin{equation}
    \frac{d}{dt}[\hat{B}(t)]^W = - \{ H^W(\mathbf{x}, \mathbf{p}), [\hat{B}(t)]^W(\mathbf{x}, \mathbf{p}) \}_{PB},
\end{equation}
given by Hamilton's equation \cite{Imre1967, Hillery1984a}. Because the Wigner-transformed Hamiltonians that arise from the bosonic representation, Eq.~(\ref{eq:general-quadratic-fermion-hamiltonian-in-cartesian-operators}), are quadratic, classical evolution of the Cartesian variables is sufficient to capture the exact dynamics of the original quantum mechanical problem, as has been observed numerically previously \cite{Li2012, Li2014b, Liu2017, Levy2019, sunBosonicPerspectiveClassical2021}. Indeed, this is also true for other mapping approaches suggested previously \cite{Li2012, Li2014b, Levy2019}, which we show in Appendix \ref{app:other-maps} can also be written in terms of quadratic bosonic operators. Thus, the analysis that we provide here for determining when a simple replacement of fermionic operators by bosonic ones can yield exact dynamics applies directly to quasiclassical mappings which can be demonstrated to behave bosonically. 

%


\begin{thebibliography}{106}%
\makeatletter
\providecommand \@ifxundefined [1]{%
 \@ifx{#1\undefined}
}%
\providecommand \@ifnum [1]{%
 \ifnum #1\expandafter \@firstoftwo
 \else \expandafter \@secondoftwo
 \fi
}%
\providecommand \@ifx [1]{%
 \ifx #1\expandafter \@firstoftwo
 \else \expandafter \@secondoftwo
 \fi
}%
\providecommand \natexlab [1]{#1}%
\providecommand \enquote  [1]{``#1''}%
\providecommand \bibnamefont  [1]{#1}%
\providecommand \bibfnamefont [1]{#1}%
\providecommand \citenamefont [1]{#1}%
\providecommand \href@noop [0]{\@secondoftwo}%
\providecommand \href [0]{\begingroup \@sanitize@url \@href}%
\providecommand \@href[1]{\@@startlink{#1}\@@href}%
\providecommand \@@href[1]{\endgroup#1\@@endlink}%
\providecommand \@sanitize@url [0]{\catcode `\\12\catcode `\$12\catcode
  `\&12\catcode `\#12\catcode `\^12\catcode `\_12\catcode `\%12\relax}%
\providecommand \@@startlink[1]{}%
\providecommand \@@endlink[0]{}%
\providecommand \url  [0]{\begingroup\@sanitize@url \@url }%
\providecommand \@url [1]{\endgroup\@href {#1}{\urlprefix }}%
\providecommand \urlprefix  [0]{URL }%
\providecommand \Eprint [0]{\href }%
\providecommand \doibase [0]{http://dx.doi.org/}%
\providecommand \selectlanguage [0]{\@gobble}%
\providecommand \bibinfo  [0]{\@secondoftwo}%
\providecommand \bibfield  [0]{\@secondoftwo}%
\providecommand \translation [1]{[#1]}%
\providecommand \BibitemOpen [0]{}%
\providecommand \bibitemStop [0]{}%
\providecommand \bibitemNoStop [0]{.\EOS\space}%
\providecommand \EOS [0]{\spacefactor3000\relax}%
\providecommand \BibitemShut  [1]{\csname bibitem#1\endcsname}%
\let\auto@bib@innerbib\@empty
\bibitem [{\citenamefont {Auerbach}(1998)}]{Auerbach1998}%
  \BibitemOpen
  \bibfield  {author} {\bibinfo {author} {\bibfnamefont {A.}~\bibnamefont
  {Auerbach}},\ }\href@noop {} {\emph {\bibinfo {title} {{Interacting Electrons
  and Quantum Magnetism}}}}\ (\bibinfo  {publisher} {Springer-Verlag},\
  \bibinfo {address} {New York},\ \bibinfo {year} {1998})\BibitemShut {NoStop}%
\bibitem [{\citenamefont
  {Dagotto}(1994)}]{dagottoCorrelatedElectronsHightemperature1994}%
  \BibitemOpen
  \bibfield  {author} {\bibinfo {author} {\bibfnamefont {E.}~\bibnamefont
  {Dagotto}},\ }\href {\doibase 10.1103/RevModPhys.66.763} {\bibfield
  {journal} {\bibinfo  {journal} {Reviews of Modern Physics}\ }\textbf
  {\bibinfo {volume} {66}},\ \bibinfo {pages} {763} (\bibinfo {year}
  {1994})}\BibitemShut {NoStop}%
\bibitem [{\citenamefont {Orenstein}\ and\ \citenamefont
  {Millis}(2000)}]{orensteinAdvancesPhysicsHighTemperature2000}%
  \BibitemOpen
  \bibfield  {author} {\bibinfo {author} {\bibfnamefont {J.}~\bibnamefont
  {Orenstein}}\ and\ \bibinfo {author} {\bibfnamefont {A.~J.}\ \bibnamefont
  {Millis}},\ }\href {\doibase 10.1126/science.288.5465.468} {\bibfield
  {journal} {\bibinfo  {journal} {Science}\ }\textbf {\bibinfo {volume}
  {288}},\ \bibinfo {pages} {468} (\bibinfo {year} {2000})}\BibitemShut
  {NoStop}%
\bibitem [{\citenamefont {Lee}, \citenamefont {Nagaosa},\ and\ \citenamefont
  {Wen}(2006)}]{leeDopingMottInsulator2006}%
  \BibitemOpen
  \bibfield  {author} {\bibinfo {author} {\bibfnamefont {P.~A.}\ \bibnamefont
  {Lee}}, \bibinfo {author} {\bibfnamefont {N.}~\bibnamefont {Nagaosa}}, \ and\
  \bibinfo {author} {\bibfnamefont {X.-G.}\ \bibnamefont {Wen}},\ }\href
  {\doibase 10.1103/RevModPhys.78.17} {\bibfield  {journal} {\bibinfo
  {journal} {Reviews of Modern Physics}\ }\textbf {\bibinfo {volume} {78}},\
  \bibinfo {pages} {17} (\bibinfo {year} {2006})}\BibitemShut {NoStop}%
\bibitem [{\citenamefont {Chidsey}(1991)}]{chidseyFreeEnergyTemperature1991}%
  \BibitemOpen
  \bibfield  {author} {\bibinfo {author} {\bibfnamefont {C.~E.~D.}\
  \bibnamefont {Chidsey}},\ }\href {\doibase 10.1126/science.251.4996.919}
  {\bibfield  {journal} {\bibinfo  {journal} {Science}\ }\textbf {\bibinfo
  {volume} {251}},\ \bibinfo {pages} {919} (\bibinfo {year}
  {1991})}\BibitemShut {NoStop}%
\bibitem [{\citenamefont {Seh}\ \emph {et~al.}(2017)\citenamefont {Seh},
  \citenamefont {Kibsgaard}, \citenamefont {Dickens}, \citenamefont
  {Chorkendorff}, \citenamefont {N{\o}rskov},\ and\ \citenamefont
  {Jaramillo}}]{sehCombiningTheoryExperiment2017}%
  \BibitemOpen
  \bibfield  {author} {\bibinfo {author} {\bibfnamefont {Z.~W.}\ \bibnamefont
  {Seh}}, \bibinfo {author} {\bibfnamefont {J.}~\bibnamefont {Kibsgaard}},
  \bibinfo {author} {\bibfnamefont {C.~F.}\ \bibnamefont {Dickens}}, \bibinfo
  {author} {\bibfnamefont {I.}~\bibnamefont {Chorkendorff}}, \bibinfo {author}
  {\bibfnamefont {J.~K.}\ \bibnamefont {N{\o}rskov}}, \ and\ \bibinfo {author}
  {\bibfnamefont {T.~F.}\ \bibnamefont {Jaramillo}},\ }\href {\doibase
  10.1126/science.aad4998} {\bibfield  {journal} {\bibinfo  {journal}
  {Science}\ }\textbf {\bibinfo {volume} {355}},\ \bibinfo {pages} {eaad4998}
  (\bibinfo {year} {2017})}\BibitemShut {NoStop}%
\bibitem [{\citenamefont {Warburton}, \citenamefont {Soudackov},\ and\
  \citenamefont
  {{Hammes-Schiffer}}(2022)}]{warburtonTheoreticalModelingElectrochemical2022}%
  \BibitemOpen
  \bibfield  {author} {\bibinfo {author} {\bibfnamefont {R.~E.}\ \bibnamefont
  {Warburton}}, \bibinfo {author} {\bibfnamefont {A.~V.}\ \bibnamefont
  {Soudackov}}, \ and\ \bibinfo {author} {\bibfnamefont {S.}~\bibnamefont
  {{Hammes-Schiffer}}},\ }\href {\doibase 10.1021/acs.chemrev.1c00929}
  {\bibfield  {journal} {\bibinfo  {journal} {Chemical Reviews}\ }\textbf
  {\bibinfo {volume} {122}},\ \bibinfo {pages} {10599} (\bibinfo {year}
  {2022})}\BibitemShut {NoStop}%
\bibitem [{\citenamefont {Santos}\ and\ \citenamefont
  {Schmickler}(2022)}]{santosModelsElectronTransfer2022}%
  \BibitemOpen
  \bibfield  {author} {\bibinfo {author} {\bibfnamefont {E.}~\bibnamefont
  {Santos}}\ and\ \bibinfo {author} {\bibfnamefont {W.}~\bibnamefont
  {Schmickler}},\ }\href {\doibase 10.1021/acs.chemrev.1c00583} {\bibfield
  {journal} {\bibinfo  {journal} {Chemical Reviews}\ }\textbf {\bibinfo
  {volume} {122}},\ \bibinfo {pages} {10581} (\bibinfo {year}
  {2022})}\BibitemShut {NoStop}%
\bibitem [{\citenamefont {Migliore}\ \emph {et~al.}(2014)\citenamefont
  {Migliore}, \citenamefont {Polizzi}, \citenamefont {Therien},\ and\
  \citenamefont {Beratan}}]{miglioreBiochemistryTheoryProtonCoupled2014}%
  \BibitemOpen
  \bibfield  {author} {\bibinfo {author} {\bibfnamefont {A.}~\bibnamefont
  {Migliore}}, \bibinfo {author} {\bibfnamefont {N.~F.}\ \bibnamefont
  {Polizzi}}, \bibinfo {author} {\bibfnamefont {M.~J.}\ \bibnamefont
  {Therien}}, \ and\ \bibinfo {author} {\bibfnamefont {D.~N.}\ \bibnamefont
  {Beratan}},\ }\href {\doibase 10.1021/cr4006654} {\bibfield  {journal}
  {\bibinfo  {journal} {Chemical Reviews}\ }\textbf {\bibinfo {volume} {114}},\
  \bibinfo {pages} {3381} (\bibinfo {year} {2014})}\BibitemShut {NoStop}%
\bibitem [{\citenamefont {Yuly}\ \emph {et~al.}(2019)\citenamefont {Yuly},
  \citenamefont {Lubner}, \citenamefont {Zhang}, \citenamefont {Beratan},\ and\
  \citenamefont {Peters}}]{yulyElectronBifurcationProgress2019}%
  \BibitemOpen
  \bibfield  {author} {\bibinfo {author} {\bibfnamefont {J.~L.}\ \bibnamefont
  {Yuly}}, \bibinfo {author} {\bibfnamefont {C.~E.}\ \bibnamefont {Lubner}},
  \bibinfo {author} {\bibfnamefont {P.}~\bibnamefont {Zhang}}, \bibinfo
  {author} {\bibfnamefont {D.~N.}\ \bibnamefont {Beratan}}, \ and\ \bibinfo
  {author} {\bibfnamefont {J.~W.}\ \bibnamefont {Peters}},\ }\href {\doibase
  10.1039/C9CC05611D} {\bibfield  {journal} {\bibinfo  {journal} {Chemical
  Communications}\ }\textbf {\bibinfo {volume} {55}},\ \bibinfo {pages} {11823}
  (\bibinfo {year} {2019})}\BibitemShut {NoStop}%
\bibitem [{\citenamefont {Pannwitz}\ and\ \citenamefont
  {S.~Wenger}(2019)}]{pannwitzProtoncoupledMultielectronTransfer2019}%
  \BibitemOpen
  \bibfield  {author} {\bibinfo {author} {\bibfnamefont {A.}~\bibnamefont
  {Pannwitz}}\ and\ \bibinfo {author} {\bibfnamefont {O.}~\bibnamefont
  {S.~Wenger}},\ }\href {\doibase 10.1039/C9CC00821G} {\bibfield  {journal}
  {\bibinfo  {journal} {Chemical Communications}\ }\textbf {\bibinfo {volume}
  {55}},\ \bibinfo {pages} {4004} (\bibinfo {year} {2019})}\BibitemShut
  {NoStop}%
\bibitem [{\citenamefont {Rutledge}\ and\ \citenamefont
  {Tezcan}(2020)}]{rutledgeElectronTransferNitrogenase2020}%
  \BibitemOpen
  \bibfield  {author} {\bibinfo {author} {\bibfnamefont {H.~L.}\ \bibnamefont
  {Rutledge}}\ and\ \bibinfo {author} {\bibfnamefont {F.~A.}\ \bibnamefont
  {Tezcan}},\ }\href {\doibase 10.1021/acs.chemrev.9b00663} {\bibfield
  {journal} {\bibinfo  {journal} {Chemical Reviews}\ }\textbf {\bibinfo
  {volume} {120}},\ \bibinfo {pages} {5158} (\bibinfo {year}
  {2020})}\BibitemShut {NoStop}%
\bibitem [{\citenamefont {Evers}\ \emph {et~al.}(2019)\citenamefont {Evers},
  \citenamefont {Koryt{\'{a}}r}, \citenamefont {Tewari},\ and\ \citenamefont
  {van Ruitenbeek}}]{Evers2019}%
  \BibitemOpen
  \bibfield  {author} {\bibinfo {author} {\bibfnamefont {F.}~\bibnamefont
  {Evers}}, \bibinfo {author} {\bibfnamefont {R.}~\bibnamefont
  {Koryt{\'{a}}r}}, \bibinfo {author} {\bibfnamefont {S.}~\bibnamefont
  {Tewari}}, \ and\ \bibinfo {author} {\bibfnamefont {J.~M.}\ \bibnamefont {van
  Ruitenbeek}},\ }\href {\doibase 10.1103/RevModPhys.92.035001} {\bibfield
  {journal} {\bibinfo  {journal} {Reviews of Modern Physics}\ }\textbf
  {\bibinfo {volume} {92}},\ \bibinfo {pages} {35001} (\bibinfo {year}
  {2019})},\ \Eprint {http://arxiv.org/abs/1906.10449} {arXiv:1906.10449}
  \BibitemShut {NoStop}%
\bibitem [{\citenamefont {Cohen}\ and\ \citenamefont
  {Galperin}(2020)}]{CohenGalperin2020}%
  \BibitemOpen
  \bibfield  {author} {\bibinfo {author} {\bibfnamefont {G.}~\bibnamefont
  {Cohen}}\ and\ \bibinfo {author} {\bibfnamefont {M.}~\bibnamefont
  {Galperin}},\ }\href {\doibase 10.1063/1.5145210} {\bibfield  {journal}
  {\bibinfo  {journal} {J. Chem. Phys.}\ }\textbf {\bibinfo {volume} {152}},\
  \bibinfo {pages} {090901} (\bibinfo {year} {2020})},\ \Eprint
  {http://arxiv.org/abs/2001.06008} {arXiv:2001.06008} \BibitemShut {NoStop}%
\bibitem [{\citenamefont {Atia}\ and\ \citenamefont
  {Aharonov}(2017)}]{atiaFastforwardingHamiltoniansExponentially2017}%
  \BibitemOpen
  \bibfield  {author} {\bibinfo {author} {\bibfnamefont {Y.}~\bibnamefont
  {Atia}}\ and\ \bibinfo {author} {\bibfnamefont {D.}~\bibnamefont
  {Aharonov}},\ }\href {\doibase 10.1038/s41467-017-01637-7} {\bibfield
  {journal} {\bibinfo  {journal} {Nature Communications}\ }\textbf {\bibinfo
  {volume} {8}},\ \bibinfo {pages} {1572} (\bibinfo {year} {2017})}\BibitemShut
  {NoStop}%
\bibitem [{\citenamefont {Lamm}\ and\ \citenamefont
  {Lawrence}(2018)}]{lammSimulationNonequilibriumDynamics2018}%
  \BibitemOpen
  \bibfield  {author} {\bibinfo {author} {\bibfnamefont {H.}~\bibnamefont
  {Lamm}}\ and\ \bibinfo {author} {\bibfnamefont {S.}~\bibnamefont
  {Lawrence}},\ }\href {\doibase 10.1103/PhysRevLett.121.170501} {\bibfield
  {journal} {\bibinfo  {journal} {Physical Review Letters}\ }\textbf {\bibinfo
  {volume} {121}},\ \bibinfo {pages} {170501} (\bibinfo {year}
  {2018})}\BibitemShut {NoStop}%
\bibitem [{\citenamefont {Sun}\ \emph {et~al.}(2021)\citenamefont {Sun},
  \citenamefont {Motta}, \citenamefont {Tazhigulov}, \citenamefont {Tan},
  \citenamefont {Chan},\ and\ \citenamefont
  {Minnich}}]{sunQuantumComputationFiniteTemperature2021}%
  \BibitemOpen
  \bibfield  {author} {\bibinfo {author} {\bibfnamefont {S.-N.}\ \bibnamefont
  {Sun}}, \bibinfo {author} {\bibfnamefont {M.}~\bibnamefont {Motta}}, \bibinfo
  {author} {\bibfnamefont {R.~N.}\ \bibnamefont {Tazhigulov}}, \bibinfo
  {author} {\bibfnamefont {A.~T.}\ \bibnamefont {Tan}}, \bibinfo {author}
  {\bibfnamefont {G.~K.-L.}\ \bibnamefont {Chan}}, \ and\ \bibinfo {author}
  {\bibfnamefont {A.~J.}\ \bibnamefont {Minnich}},\ }\href {\doibase
  10.1103/PRXQuantum.2.010317} {\bibfield  {journal} {\bibinfo  {journal} {PRX
  Quantum}\ }\textbf {\bibinfo {volume} {2}},\ \bibinfo {pages} {010317}
  (\bibinfo {year} {2021})}\BibitemShut {NoStop}%
\bibitem [{\citenamefont {Oftelie}\ \emph {et~al.}(2022)\citenamefont
  {Oftelie}, \citenamefont {Van~Beeumen}, \citenamefont {Younis}, \citenamefont
  {Smith}, \citenamefont {Iancu},\ and\ \citenamefont {{de
  Jong}}}]{oftelieConstantdepthCircuitsDynamic2022}%
  \BibitemOpen
  \bibfield  {author} {\bibinfo {author} {\bibfnamefont {L.~B.}\ \bibnamefont
  {Oftelie}}, \bibinfo {author} {\bibfnamefont {R.}~\bibnamefont
  {Van~Beeumen}}, \bibinfo {author} {\bibfnamefont {E.}~\bibnamefont {Younis}},
  \bibinfo {author} {\bibfnamefont {E.}~\bibnamefont {Smith}}, \bibinfo
  {author} {\bibfnamefont {C.}~\bibnamefont {Iancu}}, \ and\ \bibinfo {author}
  {\bibfnamefont {W.~A.}\ \bibnamefont {{de Jong}}},\ }\href {\doibase
  10.1186/s41313-022-00043-x} {\bibfield  {journal} {\bibinfo  {journal}
  {Materials Theory}\ }\textbf {\bibinfo {volume} {6}},\ \bibinfo {pages} {13}
  (\bibinfo {year} {2022})}\BibitemShut {NoStop}%
\bibitem [{\citenamefont {Kapral}(2015)}]{Kapral2015}%
  \BibitemOpen
  \bibfield  {author} {\bibinfo {author} {\bibfnamefont {R.}~\bibnamefont
  {Kapral}},\ }\href {\doibase 10.1088/0953-8984/27/7/073201} {\bibfield
  {journal} {\bibinfo  {journal} {J. Phys.: Condens. Matter}\ }\textbf
  {\bibinfo {volume} {27}},\ \bibinfo {pages} {073201} (\bibinfo {year}
  {2015})}\BibitemShut {NoStop}%
\bibitem [{\citenamefont {Lee}, \citenamefont {Huo},\ and\ \citenamefont
  {Coker}(2016)}]{Lee2016b}%
  \BibitemOpen
  \bibfield  {author} {\bibinfo {author} {\bibfnamefont {M.~K.}\ \bibnamefont
  {Lee}}, \bibinfo {author} {\bibfnamefont {P.}~\bibnamefont {Huo}}, \ and\
  \bibinfo {author} {\bibfnamefont {D.~F.}\ \bibnamefont {Coker}},\ }\href
  {\doibase 10.1146/annurev-physchem-040215-112252} {\bibfield  {journal}
  {\bibinfo  {journal} {Annu. Rev. Phys. Chem.}\ }\textbf {\bibinfo {volume}
  {67}},\ \bibinfo {pages} {639} (\bibinfo {year} {2016})}\BibitemShut
  {NoStop}%
\bibitem [{\citenamefont {{Wolfgang P.
  Schleich}}(2001)}]{QuantumOpticsInPhaseSpace}%
  \BibitemOpen
  \bibfield  {author} {\bibinfo {author} {\bibnamefont {{Wolfgang P.
  Schleich}}},\ }\href@noop {} {\emph {\bibinfo {title} {{Quantum Optics in
  Phase Space}}}}\ (\bibinfo  {publisher} {Wiley-VCH Verlag},\ \bibinfo
  {address} {Berlin},\ \bibinfo {year} {2001})\BibitemShut {NoStop}%
\bibitem [{\citenamefont {Gardiner}\ and\ \citenamefont
  {Zoller}(2014)}]{gardinerQuantumWorldUltraCold2014}%
  \BibitemOpen
  \bibfield  {author} {\bibinfo {author} {\bibfnamefont {C.}~\bibnamefont
  {Gardiner}}\ and\ \bibinfo {author} {\bibfnamefont {P.}~\bibnamefont
  {Zoller}},\ }\href {\doibase 10.1142/p941} {\emph {\bibinfo {title} {The
  {{Quantum World}} of {{Ultra-Cold Atoms}} and {{Light Book I}}:
  {{Foundations}} of {{Quantum Optics}}}}},\ \bibinfo {series} {Cold
  {{Atoms}}}, Vol.~\bibinfo {volume} {2}\ (\bibinfo  {publisher} {{IMPERIAL
  COLLEGE PRESS}},\ \bibinfo {year} {2014})\BibitemShut {NoStop}%
\bibitem [{\citenamefont {Polkovnikov}(2010)}]{Polkovnikov2010}%
  \BibitemOpen
  \bibfield  {author} {\bibinfo {author} {\bibfnamefont {A.}~\bibnamefont
  {Polkovnikov}},\ }\href {\doibase 10.1016/j.aop.2010.02.006} {\bibfield
  {journal} {\bibinfo  {journal} {Ann. Phys.}\ }\textbf {\bibinfo {volume}
  {325}},\ \bibinfo {pages} {1790} (\bibinfo {year} {2010})}\BibitemShut
  {NoStop}%
\bibitem [{\citenamefont {Schachenmayer}, \citenamefont {Pikovski},\ and\
  \citenamefont {Rey}(2015)}]{Schachenmayer2015a}%
  \BibitemOpen
  \bibfield  {author} {\bibinfo {author} {\bibfnamefont {J.}~\bibnamefont
  {Schachenmayer}}, \bibinfo {author} {\bibfnamefont {A.}~\bibnamefont
  {Pikovski}}, \ and\ \bibinfo {author} {\bibfnamefont {A.~M.}\ \bibnamefont
  {Rey}},\ }\href {\doibase 10.1103/PhysRevX.5.011022} {\bibfield  {journal}
  {\bibinfo  {journal} {Phys. Rev. X}\ }\textbf {\bibinfo {volume} {5}},\
  \bibinfo {pages} {11022} (\bibinfo {year} {2015})}\BibitemShut {NoStop}%
\bibitem [{\citenamefont {Swingle}(2018)}]{Swingle2018}%
  \BibitemOpen
  \bibfield  {author} {\bibinfo {author} {\bibfnamefont {B.}~\bibnamefont
  {Swingle}},\ }\href {\doibase 10.1038/s41567-018-0295-5} {\bibfield
  {journal} {\bibinfo  {journal} {Nature Physics}\ }\textbf {\bibinfo {volume}
  {14}},\ \bibinfo {pages} {988} (\bibinfo {year} {2018})}\BibitemShut
  {NoStop}%
\bibitem [{\citenamefont {Montoya-Castillo}\ and\ \citenamefont
  {Markland}(2018)}]{Montoya2018}%
  \BibitemOpen
  \bibfield  {author} {\bibinfo {author} {\bibfnamefont {A.}~\bibnamefont
  {Montoya-Castillo}}\ and\ \bibinfo {author} {\bibfnamefont {T.~E.}\
  \bibnamefont {Markland}},\ }\href {\doibase 10.1038/s41598-018-31162-6}
  {\bibfield  {journal} {\bibinfo  {journal} {Sci. Rep.}\ }\textbf {\bibinfo
  {volume} {8}},\ \bibinfo {pages} {12929} (\bibinfo {year} {2018})},\ \Eprint
  {http://arxiv.org/abs/1803.05561} {arXiv:1803.05561} \BibitemShut {NoStop}%
\bibitem [{\citenamefont {Li}\ and\ \citenamefont {Miller}(2012)}]{Li2012}%
  \BibitemOpen
  \bibfield  {author} {\bibinfo {author} {\bibfnamefont {B.}~\bibnamefont
  {Li}}\ and\ \bibinfo {author} {\bibfnamefont {W.~H.}\ \bibnamefont
  {Miller}},\ }\href {\doibase 10.1063/1.4757935} {\bibfield  {journal}
  {\bibinfo  {journal} {J. Chem. Phys.}\ }\textbf {\bibinfo {volume} {137}},\
  \bibinfo {pages} {154107} (\bibinfo {year} {2012})}\BibitemShut {NoStop}%
\bibitem [{\citenamefont {Li}\ \emph {et~al.}(2014{\natexlab{a}})\citenamefont
  {Li}, \citenamefont {Miller}, \citenamefont {Levy},\ and\ \citenamefont
  {Rabani}}]{Li2014b}%
  \BibitemOpen
  \bibfield  {author} {\bibinfo {author} {\bibfnamefont {B.}~\bibnamefont
  {Li}}, \bibinfo {author} {\bibfnamefont {W.~H.}\ \bibnamefont {Miller}},
  \bibinfo {author} {\bibfnamefont {T.~J.}\ \bibnamefont {Levy}}, \ and\
  \bibinfo {author} {\bibfnamefont {E.}~\bibnamefont {Rabani}},\ }\href
  {\doibase 10.1063/1.4878736} {\bibfield  {journal} {\bibinfo  {journal} {J.
  Chem. Phys.}\ }\textbf {\bibinfo {volume} {140}},\ \bibinfo {pages} {204106}
  (\bibinfo {year} {2014}{\natexlab{a}})}\BibitemShut {NoStop}%
\bibitem [{\citenamefont {Levy}\ \emph {et~al.}(2019)\citenamefont {Levy},
  \citenamefont {Dou}, \citenamefont {Rabani},\ and\ \citenamefont
  {Limmer}}]{Levy2019}%
  \BibitemOpen
  \bibfield  {author} {\bibinfo {author} {\bibfnamefont {A.}~\bibnamefont
  {Levy}}, \bibinfo {author} {\bibfnamefont {W.}~\bibnamefont {Dou}}, \bibinfo
  {author} {\bibfnamefont {E.}~\bibnamefont {Rabani}}, \ and\ \bibinfo {author}
  {\bibfnamefont {D.~T.}\ \bibnamefont {Limmer}},\ }\href {\doibase
  10.1063/1.5099987} {\bibfield  {journal} {\bibinfo  {journal} {J. Chem.
  Phys.}\ }\textbf {\bibinfo {volume} {150}},\ \bibinfo {pages} {234112}
  (\bibinfo {year} {2019})}\BibitemShut {NoStop}%
\bibitem [{\citenamefont {Sun}\ and\ \citenamefont
  {Miller}(1997{\natexlab{a}})}]{Sun1997}%
  \BibitemOpen
  \bibfield  {author} {\bibinfo {author} {\bibfnamefont {X.}~\bibnamefont
  {Sun}}\ and\ \bibinfo {author} {\bibfnamefont {W.~H.}\ \bibnamefont
  {Miller}},\ }\href {\doibase 10.1063/1.473171} {\bibfield  {journal}
  {\bibinfo  {journal} {J. Chem. Phys.}\ }\textbf {\bibinfo {volume} {106}},\
  \bibinfo {pages} {916} (\bibinfo {year} {1997}{\natexlab{a}})}\BibitemShut
  {NoStop}%
\bibitem [{\citenamefont {M{\"{u}}ller}\ and\ \citenamefont
  {Stock}(1998)}]{Muller1998}%
  \BibitemOpen
  \bibfield  {author} {\bibinfo {author} {\bibfnamefont {U.}~\bibnamefont
  {M{\"{u}}ller}}\ and\ \bibinfo {author} {\bibfnamefont {G.}~\bibnamefont
  {Stock}},\ }\href {\doibase 10.1063/1.476184} {\bibfield  {journal} {\bibinfo
   {journal} {J. Chem. Phys.}\ }\textbf {\bibinfo {volume} {108}},\ \bibinfo
  {pages} {7516} (\bibinfo {year} {1998})}\BibitemShut {NoStop}%
\bibitem [{\citenamefont {M{\"{u}}ller}\ and\ \citenamefont
  {Stock}(1999)}]{Muller1999}%
  \BibitemOpen
  \bibfield  {author} {\bibinfo {author} {\bibfnamefont {U.}~\bibnamefont
  {M{\"{u}}ller}}\ and\ \bibinfo {author} {\bibfnamefont {G.}~\bibnamefont
  {Stock}},\ }\href {\doibase 10.1063/1.479255} {\bibfield  {journal} {\bibinfo
   {journal} {J. Chem. Phys.}\ }\textbf {\bibinfo {volume} {111}},\ \bibinfo
  {pages} {65} (\bibinfo {year} {1999})}\BibitemShut {NoStop}%
\bibitem [{\citenamefont {Wang}, \citenamefont {Sun},\ and\ \citenamefont
  {Miller}(1998)}]{Wang1998b}%
  \BibitemOpen
  \bibfield  {author} {\bibinfo {author} {\bibfnamefont {H.}~\bibnamefont
  {Wang}}, \bibinfo {author} {\bibfnamefont {X.}~\bibnamefont {Sun}}, \ and\
  \bibinfo {author} {\bibfnamefont {W.~H.}\ \bibnamefont {Miller}},\ }\href
  {\doibase 10.1063/1.476447} {\bibfield  {journal} {\bibinfo  {journal} {J.
  Chem. Phys.}\ }\textbf {\bibinfo {volume} {108}},\ \bibinfo {pages} {9726}
  (\bibinfo {year} {1998})}\BibitemShut {NoStop}%
\bibitem [{\citenamefont {Sun}, \citenamefont {Wang},\ and\ \citenamefont
  {Miller}(1998)}]{Sun1998}%
  \BibitemOpen
  \bibfield  {author} {\bibinfo {author} {\bibfnamefont {X.}~\bibnamefont
  {Sun}}, \bibinfo {author} {\bibfnamefont {H.}~\bibnamefont {Wang}}, \ and\
  \bibinfo {author} {\bibfnamefont {W.~H.}\ \bibnamefont {Miller}},\ }\href
  {\doibase 10.1063/1.477389} {\bibfield  {journal} {\bibinfo  {journal} {J.
  Chem. Phys.}\ }\textbf {\bibinfo {volume} {109}},\ \bibinfo {pages} {7064}
  (\bibinfo {year} {1998})}\BibitemShut {NoStop}%
\bibitem [{\citenamefont {Thoss}\ and\ \citenamefont
  {Stock}(1999)}]{Thoss1999}%
  \BibitemOpen
  \bibfield  {author} {\bibinfo {author} {\bibfnamefont {M.}~\bibnamefont
  {Thoss}}\ and\ \bibinfo {author} {\bibfnamefont {G.}~\bibnamefont {Stock}},\
  }\href {\doibase 10.1103/PhysRevA.59.64} {\bibfield  {journal} {\bibinfo
  {journal} {Phys. Rev. A}\ }\textbf {\bibinfo {volume} {59}},\ \bibinfo
  {pages} {64} (\bibinfo {year} {1999})}\BibitemShut {NoStop}%
\bibitem [{\citenamefont {Thoss}, \citenamefont {Miller},\ and\ \citenamefont
  {Stock}(2000)}]{Thoss2000}%
  \BibitemOpen
  \bibfield  {author} {\bibinfo {author} {\bibfnamefont {M.}~\bibnamefont
  {Thoss}}, \bibinfo {author} {\bibfnamefont {W.~H.}\ \bibnamefont {Miller}}, \
  and\ \bibinfo {author} {\bibfnamefont {G.}~\bibnamefont {Stock}},\ }\href
  {\doibase 10.1063/1.481668} {\bibfield  {journal} {\bibinfo  {journal} {J.
  Chem. Phys.}\ }\textbf {\bibinfo {volume} {112}},\ \bibinfo {pages} {10282}
  (\bibinfo {year} {2000})}\BibitemShut {NoStop}%
\bibitem [{\citenamefont {Volobuev}\ \emph {et~al.}(2000)\citenamefont
  {Volobuev}, \citenamefont {Hack}, \citenamefont {Topaler},\ and\
  \citenamefont {Truhlar}}]{Volobuev2000}%
  \BibitemOpen
  \bibfield  {author} {\bibinfo {author} {\bibfnamefont {Y.~L.}\ \bibnamefont
  {Volobuev}}, \bibinfo {author} {\bibfnamefont {M.~D.}\ \bibnamefont {Hack}},
  \bibinfo {author} {\bibfnamefont {M.~S.}\ \bibnamefont {Topaler}}, \ and\
  \bibinfo {author} {\bibfnamefont {D.~G.}\ \bibnamefont {Truhlar}},\ }\href
  {\doibase 10.1063/1.481609} {\bibfield  {journal} {\bibinfo  {journal} {J.
  Chem. Phys.}\ }\textbf {\bibinfo {volume} {112}},\ \bibinfo {pages} {9716}
  (\bibinfo {year} {2000})}\BibitemShut {NoStop}%
\bibitem [{\citenamefont {Coronado}, \citenamefont {Xing},\ and\ \citenamefont
  {Miller}(2001)}]{Coronado2001}%
  \BibitemOpen
  \bibfield  {author} {\bibinfo {author} {\bibfnamefont {E.~A.}\ \bibnamefont
  {Coronado}}, \bibinfo {author} {\bibfnamefont {J.}~\bibnamefont {Xing}}, \
  and\ \bibinfo {author} {\bibfnamefont {W.~H.}\ \bibnamefont {Miller}},\
  }\href {\doibase 10.1016/S0009-2614(01)01242-8} {\bibfield  {journal}
  {\bibinfo  {journal} {Chem. Phys. Lett.}\ }\textbf {\bibinfo {volume}
  {349}},\ \bibinfo {pages} {521} (\bibinfo {year} {2001})}\BibitemShut
  {NoStop}%
\bibitem [{\citenamefont {Liao}\ and\ \citenamefont {Voth}(2002)}]{Liao2002}%
  \BibitemOpen
  \bibfield  {author} {\bibinfo {author} {\bibfnamefont {J.-L.}\ \bibnamefont
  {Liao}}\ and\ \bibinfo {author} {\bibfnamefont {G.~A.}\ \bibnamefont
  {Voth}},\ }\href {\doibase 10.1021/jp020978d} {\bibfield  {journal} {\bibinfo
   {journal} {J. Phys. Chem. B}\ }\textbf {\bibinfo {volume} {106}},\ \bibinfo
  {pages} {8449} (\bibinfo {year} {2002})}\BibitemShut {NoStop}%
\bibitem [{\citenamefont {Shi}\ and\ \citenamefont
  {Geva}(2004{\natexlab{a}})}]{Shi2004a}%
  \BibitemOpen
  \bibfield  {author} {\bibinfo {author} {\bibfnamefont {Q.}~\bibnamefont
  {Shi}}\ and\ \bibinfo {author} {\bibfnamefont {E.}~\bibnamefont {Geva}},\
  }\href {\doibase 10.1063/1.1738109} {\bibfield  {journal} {\bibinfo
  {journal} {J. Chem. Phys.}\ }\textbf {\bibinfo {volume} {120}},\ \bibinfo
  {pages} {10647} (\bibinfo {year} {2004}{\natexlab{a}})}\BibitemShut {NoStop}%
\bibitem [{\citenamefont {Bonella}\ and\ \citenamefont
  {Coker}(2005)}]{Bonella2005}%
  \BibitemOpen
  \bibfield  {author} {\bibinfo {author} {\bibfnamefont {S.}~\bibnamefont
  {Bonella}}\ and\ \bibinfo {author} {\bibfnamefont {D.~F.}\ \bibnamefont
  {Coker}},\ }\href {\doibase 10.1063/1.1896948} {\bibfield  {journal}
  {\bibinfo  {journal} {J. Chem. Phys.}\ }\textbf {\bibinfo {volume} {122}},\
  \bibinfo {pages} {194102} (\bibinfo {year} {2005})}\BibitemShut {NoStop}%
\bibitem [{\citenamefont {Ananth}, \citenamefont {Venkataraman},\ and\
  \citenamefont {Miller}(2007)}]{Ananth2007}%
  \BibitemOpen
  \bibfield  {author} {\bibinfo {author} {\bibfnamefont {N.}~\bibnamefont
  {Ananth}}, \bibinfo {author} {\bibfnamefont {C.}~\bibnamefont
  {Venkataraman}}, \ and\ \bibinfo {author} {\bibfnamefont {W.~H.}\
  \bibnamefont {Miller}},\ }\href {\doibase 10.1063/1.2759932} {\bibfield
  {journal} {\bibinfo  {journal} {J. Chem. Phys.}\ }\textbf {\bibinfo {volume}
  {127}},\ \bibinfo {pages} {084114} (\bibinfo {year} {2007})}\BibitemShut
  {NoStop}%
\bibitem [{\citenamefont {Dunkel}, \citenamefont {Bonella},\ and\ \citenamefont
  {Coker}(2008)}]{Dunkel2008}%
  \BibitemOpen
  \bibfield  {author} {\bibinfo {author} {\bibfnamefont {E.~R.}\ \bibnamefont
  {Dunkel}}, \bibinfo {author} {\bibfnamefont {S.}~\bibnamefont {Bonella}}, \
  and\ \bibinfo {author} {\bibfnamefont {D.~F.}\ \bibnamefont {Coker}},\ }\href
  {\doibase 10.1063/1.2976441} {\bibfield  {journal} {\bibinfo  {journal} {J.
  Chem. Phys.}\ }\textbf {\bibinfo {volume} {129}},\ \bibinfo {pages} {114106}
  (\bibinfo {year} {2008})}\BibitemShut {NoStop}%
\bibitem [{\citenamefont {Kim}, \citenamefont {Nassimi},\ and\ \citenamefont
  {Kapral}(2008)}]{Kim2008}%
  \BibitemOpen
  \bibfield  {author} {\bibinfo {author} {\bibfnamefont {H.}~\bibnamefont
  {Kim}}, \bibinfo {author} {\bibfnamefont {A.}~\bibnamefont {Nassimi}}, \ and\
  \bibinfo {author} {\bibfnamefont {R.}~\bibnamefont {Kapral}},\ }\href
  {\doibase 10.1063/1.2971041} {\bibfield  {journal} {\bibinfo  {journal} {J.
  Chem. Phys.}\ }\textbf {\bibinfo {volume} {129}},\ \bibinfo {pages} {84102}
  (\bibinfo {year} {2008})}\BibitemShut {NoStop}%
\bibitem [{\citenamefont {Ananth}\ and\ \citenamefont
  {Miller}(2010)}]{Ananth2010}%
  \BibitemOpen
  \bibfield  {author} {\bibinfo {author} {\bibfnamefont {N.}~\bibnamefont
  {Ananth}}\ and\ \bibinfo {author} {\bibfnamefont {T.~F.}\ \bibnamefont
  {Miller}},\ }\href {\doibase 10.1063/1.3511700} {\bibfield  {journal}
  {\bibinfo  {journal} {J. Chem. Phys.}\ }\textbf {\bibinfo {volume} {133}},\
  \bibinfo {pages} {234103} (\bibinfo {year} {2010})}\BibitemShut {NoStop}%
\bibitem [{\citenamefont {Huo}\ and\ \citenamefont {Coker}(2011)}]{Huo2011}%
  \BibitemOpen
  \bibfield  {author} {\bibinfo {author} {\bibfnamefont {P.}~\bibnamefont
  {Huo}}\ and\ \bibinfo {author} {\bibfnamefont {D.~F.}\ \bibnamefont
  {Coker}},\ }\href {\doibase 10.1063/1.3664763} {\bibfield  {journal}
  {\bibinfo  {journal} {J. Chem. Phys.}\ }\textbf {\bibinfo {volume} {135}},\
  \bibinfo {pages} {201101} (\bibinfo {year} {2011})}\BibitemShut {NoStop}%
\bibitem [{\citenamefont {Hsieh}\ and\ \citenamefont
  {Kapral}(2012)}]{Hsieh2012}%
  \BibitemOpen
  \bibfield  {author} {\bibinfo {author} {\bibfnamefont {C.~Y.}\ \bibnamefont
  {Hsieh}}\ and\ \bibinfo {author} {\bibfnamefont {R.}~\bibnamefont {Kapral}},\
  }\href {\doibase 10.1063/1.4736841} {\bibfield  {journal} {\bibinfo
  {journal} {J. Chem. Phys.}\ }\textbf {\bibinfo {volume} {137}},\ \bibinfo
  {pages} {22A507} (\bibinfo {year} {2012})}\BibitemShut {NoStop}%
\bibitem [{\citenamefont {Kelly}\ \emph {et~al.}(2012)\citenamefont {Kelly},
  \citenamefont {van Zon}, \citenamefont {Schofield},\ and\ \citenamefont
  {Kapral}}]{Kelly2012}%
  \BibitemOpen
  \bibfield  {author} {\bibinfo {author} {\bibfnamefont {A.}~\bibnamefont
  {Kelly}}, \bibinfo {author} {\bibfnamefont {R.}~\bibnamefont {van Zon}},
  \bibinfo {author} {\bibfnamefont {J.}~\bibnamefont {Schofield}}, \ and\
  \bibinfo {author} {\bibfnamefont {R.}~\bibnamefont {Kapral}},\ }\href
  {\doibase 10.1063/1.3685420} {\bibfield  {journal} {\bibinfo  {journal} {J.
  Chem. Phys.}\ }\textbf {\bibinfo {volume} {136}},\ \bibinfo {pages} {84101}
  (\bibinfo {year} {2012})},\ \Eprint {http://arxiv.org/abs/arXiv:1201.1042v2}
  {arXiv:arXiv:1201.1042v2} \BibitemShut {NoStop}%
\bibitem [{\citenamefont {Ananth}(2013)}]{Ananth2013}%
  \BibitemOpen
  \bibfield  {author} {\bibinfo {author} {\bibfnamefont {N.}~\bibnamefont
  {Ananth}},\ }\href {\doibase 10.1063/1.4821590} {\bibfield  {journal}
  {\bibinfo  {journal} {J. Chem. Phys.}\ }\textbf {\bibinfo {volume} {139}},\
  \bibinfo {pages} {124102} (\bibinfo {year} {2013})}\BibitemShut {NoStop}%
\bibitem [{\citenamefont {Richardson}\ and\ \citenamefont
  {Thoss}(2013)}]{Richardson2013}%
  \BibitemOpen
  \bibfield  {author} {\bibinfo {author} {\bibfnamefont {J.~O.}\ \bibnamefont
  {Richardson}}\ and\ \bibinfo {author} {\bibfnamefont {M.}~\bibnamefont
  {Thoss}},\ }\href {\doibase 10.1063/1.4816124} {\bibfield  {journal}
  {\bibinfo  {journal} {J. Chem. Phys.}\ }\textbf {\bibinfo {volume} {139}},\
  \bibinfo {pages} {31102} (\bibinfo {year} {2013})}\BibitemShut {NoStop}%
\bibitem [{\citenamefont {Hele}\ and\ \citenamefont {Ananth}(2016)}]{Hele2016}%
  \BibitemOpen
  \bibfield  {author} {\bibinfo {author} {\bibfnamefont {T.~J.~H.}\
  \bibnamefont {Hele}}\ and\ \bibinfo {author} {\bibfnamefont {N.}~\bibnamefont
  {Ananth}},\ }\href {\doibase 10.1039/C6FD00106H} {\bibfield  {journal}
  {\bibinfo  {journal} {Faraday Discuss.}\ }\textbf {\bibinfo {volume} {195}},\
  \bibinfo {pages} {269} (\bibinfo {year} {2016})}\BibitemShut {NoStop}%
\bibitem [{\citenamefont {Chowdhury}\ and\ \citenamefont
  {Huo}(2017)}]{Chowdhury2017}%
  \BibitemOpen
  \bibfield  {author} {\bibinfo {author} {\bibfnamefont {S.~N.}\ \bibnamefont
  {Chowdhury}}\ and\ \bibinfo {author} {\bibfnamefont {P.}~\bibnamefont
  {Huo}},\ }\href {\doibase 10.1063/1.4995616} {\bibfield  {journal} {\bibinfo
  {journal} {J. Chem. Phys.}\ }\textbf {\bibinfo {volume} {147}},\ \bibinfo
  {pages} {214109} (\bibinfo {year} {2017})},\ \Eprint
  {http://arxiv.org/abs/1706.08403} {arXiv:1706.08403} \BibitemShut {NoStop}%
\bibitem [{\citenamefont {Church}\ \emph {et~al.}(2018)\citenamefont {Church},
  \citenamefont {Hele}, \citenamefont {Ezra},\ and\ \citenamefont
  {Ananth}}]{Church2018}%
  \BibitemOpen
  \bibfield  {author} {\bibinfo {author} {\bibfnamefont {M.~S.}\ \bibnamefont
  {Church}}, \bibinfo {author} {\bibfnamefont {T.~J.~H.}\ \bibnamefont {Hele}},
  \bibinfo {author} {\bibfnamefont {G.~S.}\ \bibnamefont {Ezra}}, \ and\
  \bibinfo {author} {\bibfnamefont {N.}~\bibnamefont {Ananth}},\ }\href
  {\doibase 10.1063/1.5005557} {\bibfield  {journal} {\bibinfo  {journal} {J.
  Chem. Phys.}\ }\textbf {\bibinfo {volume} {148}},\ \bibinfo {pages} {102326}
  (\bibinfo {year} {2018})},\ \Eprint {http://arxiv.org/abs/1709.07474}
  {arXiv:1709.07474} \BibitemShut {NoStop}%
\bibitem [{\citenamefont {Li}\ \emph {et~al.}(2013)\citenamefont {Li},
  \citenamefont {Levy}, \citenamefont {Swenson}, \citenamefont {Rabani},\ and\
  \citenamefont {Miller}}]{Li2013a}%
  \BibitemOpen
  \bibfield  {author} {\bibinfo {author} {\bibfnamefont {B.}~\bibnamefont
  {Li}}, \bibinfo {author} {\bibfnamefont {T.~J.}\ \bibnamefont {Levy}},
  \bibinfo {author} {\bibfnamefont {D.~W.~H.}\ \bibnamefont {Swenson}},
  \bibinfo {author} {\bibfnamefont {E.}~\bibnamefont {Rabani}}, \ and\ \bibinfo
  {author} {\bibfnamefont {W.~H.}\ \bibnamefont {Miller}},\ }\href {\doibase
  10.1063/1.4793747} {\bibfield  {journal} {\bibinfo  {journal} {J. Chem.
  Phys.}\ }\textbf {\bibinfo {volume} {138}},\ \bibinfo {pages} {104110}
  (\bibinfo {year} {2013})}\BibitemShut {NoStop}%
\bibitem [{\citenamefont {Li}\ \emph {et~al.}(2014{\natexlab{b}})\citenamefont
  {Li}, \citenamefont {Wilner}, \citenamefont {Thoss}, \citenamefont {Rabani},\
  and\ \citenamefont {Miller}}]{Li2014a}%
  \BibitemOpen
  \bibfield  {author} {\bibinfo {author} {\bibfnamefont {B.}~\bibnamefont
  {Li}}, \bibinfo {author} {\bibfnamefont {E.~Y.}\ \bibnamefont {Wilner}},
  \bibinfo {author} {\bibfnamefont {M.}~\bibnamefont {Thoss}}, \bibinfo
  {author} {\bibfnamefont {E.}~\bibnamefont {Rabani}}, \ and\ \bibinfo {author}
  {\bibfnamefont {W.~H.}\ \bibnamefont {Miller}},\ }\href {\doibase
  10.1063/1.4867789} {\bibfield  {journal} {\bibinfo  {journal} {J. Chem.
  Phys.}\ }\textbf {\bibinfo {volume} {140}},\ \bibinfo {pages} {104110}
  (\bibinfo {year} {2014}{\natexlab{b}})}\BibitemShut {NoStop}%
\bibitem [{\citenamefont {Liu}(2017)}]{Liu2017}%
  \BibitemOpen
  \bibfield  {author} {\bibinfo {author} {\bibfnamefont {J.}~\bibnamefont
  {Liu}},\ }\href {\doibase 10.1063/1.4973708} {\bibfield  {journal} {\bibinfo
  {journal} {J. Chem. Phys.}\ }\textbf {\bibinfo {volume} {146}},\ \bibinfo
  {pages} {024110} (\bibinfo {year} {2017})}\BibitemShut {NoStop}%
\bibitem [{\citenamefont {Sun}, \citenamefont {Sasmal},\ and\ \citenamefont
  {Vendrell}(2021)}]{sunBosonicPerspectiveClassical2021}%
  \BibitemOpen
  \bibfield  {author} {\bibinfo {author} {\bibfnamefont {J.}~\bibnamefont
  {Sun}}, \bibinfo {author} {\bibfnamefont {S.}~\bibnamefont {Sasmal}}, \ and\
  \bibinfo {author} {\bibfnamefont {O.}~\bibnamefont {Vendrell}},\ }\href
  {\doibase 10.1063/5.0066740} {\bibfield  {journal} {\bibinfo  {journal} {The
  Journal of Chemical Physics}\ }\textbf {\bibinfo {volume} {155}},\ \bibinfo
  {pages} {134110} (\bibinfo {year} {2021})}\BibitemShut {NoStop}%
\bibitem [{\citenamefont {Meyer}\ and\ \citenamefont
  {Miller}(1979)}]{Meyer1979}%
  \BibitemOpen
  \bibfield  {author} {\bibinfo {author} {\bibfnamefont {H.-D.}\ \bibnamefont
  {Meyer}}\ and\ \bibinfo {author} {\bibfnamefont {W.~H.}\ \bibnamefont
  {Miller}},\ }\href {\doibase 10.1063/1.437910} {\bibfield  {journal}
  {\bibinfo  {journal} {J. Chem. Phys.}\ }\textbf {\bibinfo {volume} {70}},\
  \bibinfo {pages} {3214} (\bibinfo {year} {1979})}\BibitemShut {NoStop}%
\bibitem [{\citenamefont {Stock}\ and\ \citenamefont
  {Thoss}(1997)}]{Stock1997}%
  \BibitemOpen
  \bibfield  {author} {\bibinfo {author} {\bibfnamefont {G.}~\bibnamefont
  {Stock}}\ and\ \bibinfo {author} {\bibfnamefont {M.}~\bibnamefont {Thoss}},\
  }\href {\doibase 10.1103/PhysRevLett.78.578} {\bibfield  {journal} {\bibinfo
  {journal} {Phys. Rev. Lett.}\ }\textbf {\bibinfo {volume} {78}},\ \bibinfo
  {pages} {578} (\bibinfo {year} {1997})}\BibitemShut {NoStop}%
\bibitem [{\citenamefont {Holstein}(1959)}]{holsteinStudiesPolaronMotion1959a}%
  \BibitemOpen
  \bibfield  {author} {\bibinfo {author} {\bibfnamefont {T.}~\bibnamefont
  {Holstein}},\ }\href {\doibase 10.1016/0003-4916(59)90002-8} {\bibfield
  {journal} {\bibinfo  {journal} {Annals of Physics}\ }\textbf {\bibinfo
  {volume} {8}},\ \bibinfo {pages} {325} (\bibinfo {year} {1959})}\BibitemShut
  {NoStop}%
\bibitem [{\citenamefont {Troisi}\ and\ \citenamefont
  {Orlandi}(2006)}]{troisiChargeTransportRegimeCrystalline2006}%
  \BibitemOpen
  \bibfield  {author} {\bibinfo {author} {\bibfnamefont {A.}~\bibnamefont
  {Troisi}}\ and\ \bibinfo {author} {\bibfnamefont {G.}~\bibnamefont
  {Orlandi}},\ }\href {\doibase 10.1103/PhysRevLett.96.086601} {\bibfield
  {journal} {\bibinfo  {journal} {Physical Review Letters}\ }\textbf {\bibinfo
  {volume} {96}},\ \bibinfo {pages} {086601} (\bibinfo {year}
  {2006})}\BibitemShut {NoStop}%
\bibitem [{\citenamefont {Fratini}, \citenamefont {Mayou},\ and\ \citenamefont
  {Ciuchi}(2016)}]{fratiniTransientLocalizationScenario2016}%
  \BibitemOpen
  \bibfield  {author} {\bibinfo {author} {\bibfnamefont {S.}~\bibnamefont
  {Fratini}}, \bibinfo {author} {\bibfnamefont {D.}~\bibnamefont {Mayou}}, \
  and\ \bibinfo {author} {\bibfnamefont {S.}~\bibnamefont {Ciuchi}},\ }\href
  {\doibase 10.1002/adfm.201502386} {\bibfield  {journal} {\bibinfo  {journal}
  {Advanced Functional Materials}\ }\textbf {\bibinfo {volume} {26}},\ \bibinfo
  {pages} {2292} (\bibinfo {year} {2016})}\BibitemShut {NoStop}%
\bibitem [{\citenamefont {Hewson}(1993)}]{Hewson1993-Kondo-book}%
  \BibitemOpen
  \bibfield  {author} {\bibinfo {author} {\bibfnamefont {A.~C.}\ \bibnamefont
  {Hewson}},\ }\href@noop {} {\emph {\bibinfo {title} {{The Kondo Problem to
  Heavy Fermions}}}}\ (\bibinfo  {publisher} {Cambridge University Press},\
  \bibinfo {address} {Cambridge},\ \bibinfo {year} {1993})\BibitemShut
  {NoStop}%
\bibitem [{\citenamefont {Averin}\ and\ \citenamefont
  {Likharev}(1986)}]{averinCoulombBlockadeSingleelectron1986}%
  \BibitemOpen
  \bibfield  {author} {\bibinfo {author} {\bibfnamefont {D.~V.}\ \bibnamefont
  {Averin}}\ and\ \bibinfo {author} {\bibfnamefont {K.~K.}\ \bibnamefont
  {Likharev}},\ }\href {\doibase 10.1007/BF00683469} {\bibfield  {journal}
  {\bibinfo  {journal} {Journal of Low Temperature Physics}\ }\textbf {\bibinfo
  {volume} {62}},\ \bibinfo {pages} {345} (\bibinfo {year} {1986})}\BibitemShut
  {NoStop}%
\bibitem [{\citenamefont
  {Beenakker}(1991)}]{beenakkerTheoryCoulombblockadeOscillations1991}%
  \BibitemOpen
  \bibfield  {author} {\bibinfo {author} {\bibfnamefont {C.~W.~J.}\
  \bibnamefont {Beenakker}},\ }\href {\doibase 10.1103/PhysRevB.44.1646}
  {\bibfield  {journal} {\bibinfo  {journal} {Physical Review B}\ }\textbf
  {\bibinfo {volume} {44}},\ \bibinfo {pages} {1646} (\bibinfo {year}
  {1991})}\BibitemShut {NoStop}%
\bibitem [{\citenamefont {Schwinger}(1965)}]{Schwinger1965}%
  \BibitemOpen
  \bibfield  {author} {\bibinfo {author} {\bibfnamefont {J.}~\bibnamefont
  {Schwinger}},\ }in\ \href {\doibase 10.2172/4389568} {\emph {\bibinfo
  {booktitle} {Quantum theory of angular momentum}}},\ \bibinfo {editor}
  {edited by\ \bibinfo {editor} {\bibfnamefont {L.~C.}\ \bibnamefont
  {Biedenharn}}\ and\ \bibinfo {editor} {\bibfnamefont {H.~V.}\ \bibnamefont
  {Dam}}}\ (\bibinfo  {publisher} {Academic Press},\ \bibinfo {address} {New
  York},\ \bibinfo {year} {1965})\ pp.\ \bibinfo {pages} {229--279}\BibitemShut
  {NoStop}%
\bibitem [{\citenamefont {Holstein}\ and\ \citenamefont
  {Primakoff}(1940)}]{Holstein1940}%
  \BibitemOpen
  \bibfield  {author} {\bibinfo {author} {\bibfnamefont {T.}~\bibnamefont
  {Holstein}}\ and\ \bibinfo {author} {\bibfnamefont {H.}~\bibnamefont
  {Primakoff}},\ }\href {\doibase 10.1103/PhysRev.58.1098} {\bibfield
  {journal} {\bibinfo  {journal} {Phys. Rev.}\ }\textbf {\bibinfo {volume}
  {58}},\ \bibinfo {pages} {1098} (\bibinfo {year} {1940})}\BibitemShut
  {NoStop}%
\bibitem [{\citenamefont {Fisher}\ \emph {et~al.}(1989)\citenamefont {Fisher},
  \citenamefont {Weichman}, \citenamefont {Grinstein},\ and\ \citenamefont
  {Fisher}}]{fisherBosonLocalizationSuperfluidinsulator1989}%
  \BibitemOpen
  \bibfield  {author} {\bibinfo {author} {\bibfnamefont {M.~P.~A.}\
  \bibnamefont {Fisher}}, \bibinfo {author} {\bibfnamefont {P.~B.}\
  \bibnamefont {Weichman}}, \bibinfo {author} {\bibfnamefont {G.}~\bibnamefont
  {Grinstein}}, \ and\ \bibinfo {author} {\bibfnamefont {D.~S.}\ \bibnamefont
  {Fisher}},\ }\href {\doibase 10.1103/PhysRevB.40.546} {\bibfield  {journal}
  {\bibinfo  {journal} {Physical Review B}\ }\textbf {\bibinfo {volume} {40}},\
  \bibinfo {pages} {546} (\bibinfo {year} {1989})}\BibitemShut {NoStop}%
\bibitem [{\citenamefont {Giamarchi}(2004)}]{Giamarchi2004}%
  \BibitemOpen
  \bibfield  {author} {\bibinfo {author} {\bibfnamefont {T.}~\bibnamefont
  {Giamarchi}},\ }\href@noop {} {\emph {\bibinfo {title} {{Quantum Physics in
  One Dimension}}}}\ (\bibinfo  {publisher} {Oxford University Press},\
  \bibinfo {address} {New York},\ \bibinfo {year} {2004})\BibitemShut {NoStop}%
\bibitem [{\citenamefont {S{\'e}n{\'e}chal}, \citenamefont {Tremblay},\ and\
  \citenamefont {Bourbonnais}()}]{senechalTheoreticalMethodsStrongly}%
  \BibitemOpen
  \bibinfo {editor} {\bibfnamefont {D.}~\bibnamefont {S{\'e}n{\'e}chal}},
  \bibinfo {editor} {\bibfnamefont {A.-M.}\ \bibnamefont {Tremblay}}, \ and\
  \bibinfo {editor} {\bibfnamefont {C.}~\bibnamefont {Bourbonnais}},\ eds.,\
  \href@noop {} {\emph {\bibinfo {title} {Theoretical {{Methods}} for
  {{Strongly Correlated Electrons}}}}},\ {{CRM Series}} in {{Mathematical
  Physics}}\BibitemShut {NoStop}%
\bibitem [{\citenamefont {S{\'{e}}n{\'{e}}chal}, \citenamefont {Tremblay},\
  and\ \citenamefont
  {Bourbonnais}(2004)}]{Senechal2004-Theoretical-Methods-for-Strongly-Correlated-Electrons}%
  \BibitemOpen
  \bibinfo {editor} {\bibfnamefont {D.}~\bibnamefont {S{\'{e}}n{\'{e}}chal}},
  \bibinfo {editor} {\bibfnamefont {A.-M.}\ \bibnamefont {Tremblay}}, \ and\
  \bibinfo {editor} {\bibfnamefont {C.}~\bibnamefont {Bourbonnais}},\ eds.,\
  \href@noop {} {\emph {\bibinfo {title} {{Theoretical Methods for Strongly
  Correlated Electrons}}}},\ The CRM series in mathematical physics\ (\bibinfo
  {publisher} {Springer-Verlag},\ \bibinfo {address} {New York},\ \bibinfo
  {year} {2004})\BibitemShut {NoStop}%
\bibitem [{\citenamefont {Kornyshev}\ and\ \citenamefont
  {Schmickler}(1985)}]{Kornyshev1985}%
  \BibitemOpen
  \bibfield  {author} {\bibinfo {author} {\bibfnamefont {A.~A.}\ \bibnamefont
  {Kornyshev}}\ and\ \bibinfo {author} {\bibfnamefont {W.}~\bibnamefont
  {Schmickler}},\ }\href {\doibase 10.1016/0368-1874(85)80133-7} {\bibfield
  {journal} {\bibinfo  {journal} {J. Electroanal. Chem.}\ }\textbf {\bibinfo
  {volume} {185}},\ \bibinfo {pages} {253} (\bibinfo {year}
  {1985})}\BibitemShut {NoStop}%
\bibitem [{\citenamefont {Sebastian}(1989)}]{Sebastian1989}%
  \BibitemOpen
  \bibfield  {author} {\bibinfo {author} {\bibfnamefont {K.~L.}\ \bibnamefont
  {Sebastian}},\ }\href {\doibase 10.1063/1.456547} {\bibfield  {journal}
  {\bibinfo  {journal} {J. Chem. Phys.}\ }\textbf {\bibinfo {volume} {90}},\
  \bibinfo {pages} {5056} (\bibinfo {year} {1989})}\BibitemShut {NoStop}%
\bibitem [{\citenamefont {N{\o}rskov}(1990)}]{Norsvok1990}%
  \BibitemOpen
  \bibfield  {author} {\bibinfo {author} {\bibfnamefont {J.~K.}\ \bibnamefont
  {N{\o}rskov}},\ }\href {\doibase 10.1088/0034-4885/53/10/001} {\bibfield
  {journal} {\bibinfo  {journal} {Rep. Prog. Phys.}\ }\textbf {\bibinfo
  {volume} {53}},\ \bibinfo {pages} {1253} (\bibinfo {year}
  {1990})}\BibitemShut {NoStop}%
\bibitem [{\citenamefont {Kondov}\ \emph {et~al.}(2007)\citenamefont {Kondov},
  \citenamefont {C{\'{i}}{\v{z}}ek}, \citenamefont {Benesch}, \citenamefont
  {Wang},\ and\ \citenamefont {Thoss}}]{Kondov2007}%
  \BibitemOpen
  \bibfield  {author} {\bibinfo {author} {\bibfnamefont {I.}~\bibnamefont
  {Kondov}}, \bibinfo {author} {\bibfnamefont {M.}~\bibnamefont
  {C{\'{i}}{\v{z}}ek}}, \bibinfo {author} {\bibfnamefont {C.}~\bibnamefont
  {Benesch}}, \bibinfo {author} {\bibfnamefont {H.}~\bibnamefont {Wang}}, \
  and\ \bibinfo {author} {\bibfnamefont {M.}~\bibnamefont {Thoss}},\ }\href
  {\doibase 10.1021/jp072217m} {\bibfield  {journal} {\bibinfo  {journal} {J.
  Phys. Chem. C}\ }\textbf {\bibinfo {volume} {111}},\ \bibinfo {pages} {11970}
  (\bibinfo {year} {2007})}\BibitemShut {NoStop}%
\bibitem [{\citenamefont {Miller}(2001)}]{Miller2001a}%
  \BibitemOpen
  \bibfield  {author} {\bibinfo {author} {\bibfnamefont {W.~H.}\ \bibnamefont
  {Miller}},\ }\href {\doibase 10.1080/14427591.2008.9686601} {\bibfield
  {journal} {\bibinfo  {journal} {J. Phys. Chem. A}\ }\textbf {\bibinfo
  {volume} {105}},\ \bibinfo {pages} {2942} (\bibinfo {year}
  {2001})}\BibitemShut {NoStop}%
\bibitem [{\citenamefont {Stock}\ and\ \citenamefont
  {Thoss}(2005)}]{Stock2005}%
  \BibitemOpen
  \bibfield  {author} {\bibinfo {author} {\bibfnamefont {G.}~\bibnamefont
  {Stock}}\ and\ \bibinfo {author} {\bibfnamefont {M.}~\bibnamefont {Thoss}},\
  }in\ \href@noop {} {\emph {\bibinfo {booktitle} {Adv. Chem. Phys.}}},\ Vol.\
  \bibinfo {volume} {131}\ (\bibinfo {year} {2005})\ p.\ \bibinfo {pages}
  {243}\BibitemShut {NoStop}%
\bibitem [{\citenamefont {Crespo-Otero}\ and\ \citenamefont
  {Barbatti}(2018)}]{Crespo-Otero2018}%
  \BibitemOpen
  \bibfield  {author} {\bibinfo {author} {\bibfnamefont {R.}~\bibnamefont
  {Crespo-Otero}}\ and\ \bibinfo {author} {\bibfnamefont {M.}~\bibnamefont
  {Barbatti}},\ }\href {\doibase 10.1021/acs.chemrev.7b00577} {\bibfield
  {journal} {\bibinfo  {journal} {Chem. Rev.}\ }\textbf {\bibinfo {volume}
  {118}},\ \bibinfo {pages} {7026} (\bibinfo {year} {2018})}\BibitemShut
  {NoStop}%
\bibitem [{\citenamefont {Bonnet}(2020)}]{Bonnet2020}%
  \BibitemOpen
  \bibfield  {author} {\bibinfo {author} {\bibfnamefont {L.}~\bibnamefont
  {Bonnet}},\ }\href {\doibase 10.1063/5.0023137} {\bibfield  {journal}
  {\bibinfo  {journal} {Journal of Chemical Physics}\ }\textbf {\bibinfo
  {volume} {153}} (\bibinfo {year} {2020}),\ 10.1063/5.0023137}\BibitemShut
  {NoStop}%
\bibitem [{\citenamefont {McLachlan}(1964)}]{McLachlan1964}%
  \BibitemOpen
  \bibfield  {author} {\bibinfo {author} {\bibfnamefont {A.~D.}\ \bibnamefont
  {McLachlan}},\ }\href {\doibase 10.1080/00268976400100041} {\bibfield
  {journal} {\bibinfo  {journal} {Mol. Phys.}\ }\textbf {\bibinfo {volume}
  {8}},\ \bibinfo {pages} {39} (\bibinfo {year} {1964})}\BibitemShut {NoStop}%
\bibitem [{\citenamefont {Stock}(1995)}]{Stock1995}%
  \BibitemOpen
  \bibfield  {author} {\bibinfo {author} {\bibfnamefont {G.}~\bibnamefont
  {Stock}},\ }\href {\doibase 10.1063/1.469778} {\bibfield  {journal} {\bibinfo
   {journal} {J. Chem. Phys.}\ }\textbf {\bibinfo {volume} {103}},\ \bibinfo
  {pages} {1561} (\bibinfo {year} {1995})}\BibitemShut {NoStop}%
\bibitem [{\citenamefont {Sun}\ and\ \citenamefont
  {Miller}(1997{\natexlab{b}})}]{Sun1997a}%
  \BibitemOpen
  \bibfield  {author} {\bibinfo {author} {\bibfnamefont {X.}~\bibnamefont
  {Sun}}\ and\ \bibinfo {author} {\bibfnamefont {W.~H.}\ \bibnamefont
  {Miller}},\ }\href {\doibase 10.1063/1.452435} {\bibfield  {journal}
  {\bibinfo  {journal} {J. Chem. Phys.}\ }\textbf {\bibinfo {volume} {106}},\
  \bibinfo {pages} {6346} (\bibinfo {year} {1997}{\natexlab{b}})}\BibitemShut
  {NoStop}%
\bibitem [{\citenamefont {Shi}\ and\ \citenamefont
  {Geva}(2004{\natexlab{b}})}]{Shi2004}%
  \BibitemOpen
  \bibfield  {author} {\bibinfo {author} {\bibfnamefont {Q.}~\bibnamefont
  {Shi}}\ and\ \bibinfo {author} {\bibfnamefont {E.}~\bibnamefont {Geva}},\
  }\href {\doibase 10.1063/1.1771641} {\bibfield  {journal} {\bibinfo
  {journal} {J. Chem. Phys.}\ }\textbf {\bibinfo {volume} {121}},\ \bibinfo
  {pages} {3393} (\bibinfo {year} {2004}{\natexlab{b}})}\BibitemShut {NoStop}%
\bibitem [{\citenamefont {Cotton}\ and\ \citenamefont
  {Miller}(2013)}]{Cotton2013}%
  \BibitemOpen
  \bibfield  {author} {\bibinfo {author} {\bibfnamefont {S.~J.}\ \bibnamefont
  {Cotton}}\ and\ \bibinfo {author} {\bibfnamefont {W.~H.}\ \bibnamefont
  {Miller}},\ }\href {\doibase 10.1021/jp401078u} {\bibfield  {journal}
  {\bibinfo  {journal} {J. Phys. Chem.}\ }\textbf {\bibinfo {volume} {139}},\
  \bibinfo {pages} {234112} (\bibinfo {year} {2013})}\BibitemShut {NoStop}%
\bibitem [{\citenamefont {Huo}\ and\ \citenamefont {Coker}(2012)}]{Huo2012}%
  \BibitemOpen
  \bibfield  {author} {\bibinfo {author} {\bibfnamefont {P.}~\bibnamefont
  {Huo}}\ and\ \bibinfo {author} {\bibfnamefont {D.~F.}\ \bibnamefont
  {Coker}},\ }\href {\doibase 10.1063/1.4748316} {\bibfield  {journal}
  {\bibinfo  {journal} {J. Chem. Phys.}\ }\textbf {\bibinfo {volume} {137}},\
  \bibinfo {pages} {22A535} (\bibinfo {year} {2012})}\BibitemShut {NoStop}%
\bibitem [{\citenamefont {Kelly}\ and\ \citenamefont
  {Markland}(2013)}]{Kelly2013}%
  \BibitemOpen
  \bibfield  {author} {\bibinfo {author} {\bibfnamefont {A.}~\bibnamefont
  {Kelly}}\ and\ \bibinfo {author} {\bibfnamefont {T.~E.}\ \bibnamefont
  {Markland}},\ }\href {\doibase 10.1063/1.4812355} {\bibfield  {journal}
  {\bibinfo  {journal} {J. Chem. Phys.}\ }\textbf {\bibinfo {volume} {139}},\
  \bibinfo {pages} {014104} (\bibinfo {year} {2013})}\BibitemShut {NoStop}%
\bibitem [{\citenamefont {Kelly}, \citenamefont {Brackbill},\ and\
  \citenamefont {Markland}(2015)}]{Kelly2015}%
  \BibitemOpen
  \bibfield  {author} {\bibinfo {author} {\bibfnamefont {A.}~\bibnamefont
  {Kelly}}, \bibinfo {author} {\bibfnamefont {N.}~\bibnamefont {Brackbill}}, \
  and\ \bibinfo {author} {\bibfnamefont {T.~E.}\ \bibnamefont {Markland}},\
  }\href {\doibase 10.1063/1.4913686} {\bibfield  {journal} {\bibinfo
  {journal} {J. Chem. Phys.}\ }\textbf {\bibinfo {volume} {142}},\ \bibinfo
  {pages} {094110} (\bibinfo {year} {2015})}\BibitemShut {NoStop}%
\bibitem [{\citenamefont {Pfalzgraff}, \citenamefont {Kelly},\ and\
  \citenamefont {Markland}(2015)}]{Pfalzgraff2015}%
  \BibitemOpen
  \bibfield  {author} {\bibinfo {author} {\bibfnamefont {W.~C.}\ \bibnamefont
  {Pfalzgraff}}, \bibinfo {author} {\bibfnamefont {A.}~\bibnamefont {Kelly}}, \
  and\ \bibinfo {author} {\bibfnamefont {T.~E.}\ \bibnamefont {Markland}},\
  }\href {\doibase 10.1021/acs.jpclett.5b02131} {\bibfield  {journal} {\bibinfo
   {journal} {J. Phys. Chem. Lett.}\ }\textbf {\bibinfo {volume} {6}},\
  \bibinfo {pages} {4743} (\bibinfo {year} {2015})}\BibitemShut {NoStop}%
\bibitem [{\citenamefont {Montoya-Castillo}\ and\ \citenamefont
  {Reichman}(2016)}]{Montoya2016a}%
  \BibitemOpen
  \bibfield  {author} {\bibinfo {author} {\bibfnamefont {A.}~\bibnamefont
  {Montoya-Castillo}}\ and\ \bibinfo {author} {\bibfnamefont {D.~R.}\
  \bibnamefont {Reichman}},\ }\href {\doibase 10.1063/1.4975388} {\bibfield
  {journal} {\bibinfo  {journal} {J. Chem. Phys.}\ }\textbf {\bibinfo {volume}
  {144}},\ \bibinfo {pages} {184104} (\bibinfo {year} {2016})}\BibitemShut
  {NoStop}%
\bibitem [{\citenamefont {Kelly}\ \emph {et~al.}(2016)\citenamefont {Kelly},
  \citenamefont {Montoya-Castillo}, \citenamefont {Wang},\ and\ \citenamefont
  {Markland}}]{KellyMontoya2016}%
  \BibitemOpen
  \bibfield  {author} {\bibinfo {author} {\bibfnamefont {A.}~\bibnamefont
  {Kelly}}, \bibinfo {author} {\bibfnamefont {A.}~\bibnamefont
  {Montoya-Castillo}}, \bibinfo {author} {\bibfnamefont {L.}~\bibnamefont
  {Wang}}, \ and\ \bibinfo {author} {\bibfnamefont {T.~E.}\ \bibnamefont
  {Markland}},\ }\href {\doibase 10.1063/1.4948612} {\bibfield  {journal}
  {\bibinfo  {journal} {J. Chem. Phys.}\ }\textbf {\bibinfo {volume} {144}},\
  \bibinfo {pages} {184105} (\bibinfo {year} {2016})}\BibitemShut {NoStop}%
\bibitem [{\citenamefont {Montoya-Castillo}\ and\ \citenamefont
  {Reichman}(2017)}]{Montoya2017}%
  \BibitemOpen
  \bibfield  {author} {\bibinfo {author} {\bibfnamefont {A.}~\bibnamefont
  {Montoya-Castillo}}\ and\ \bibinfo {author} {\bibfnamefont {D.~R.}\
  \bibnamefont {Reichman}},\ }\href {\doibase 10.1063/1.4973646} {\bibfield
  {journal} {\bibinfo  {journal} {J. Chem. Phys.}\ }\textbf {\bibinfo {volume}
  {146}},\ \bibinfo {pages} {024107} (\bibinfo {year} {2017})}\BibitemShut
  {NoStop}%
\bibitem [{\citenamefont {Pfalzgraff}\ \emph {et~al.}(2019)\citenamefont
  {Pfalzgraff}, \citenamefont {Montoya-Castillo}, \citenamefont {Kelly},\ and\
  \citenamefont {Markland}}]{Pfalzgraff2019}%
  \BibitemOpen
  \bibfield  {author} {\bibinfo {author} {\bibfnamefont {W.~C.}\ \bibnamefont
  {Pfalzgraff}}, \bibinfo {author} {\bibfnamefont {A.}~\bibnamefont
  {Montoya-Castillo}}, \bibinfo {author} {\bibfnamefont {A.}~\bibnamefont
  {Kelly}}, \ and\ \bibinfo {author} {\bibfnamefont {T.~E.}\ \bibnamefont
  {Markland}},\ }\href@noop {} {\bibfield  {journal} {\bibinfo  {journal} {J.
  Chem. Phys.}\ }\textbf {\bibinfo {volume} {150}},\ \bibinfo {pages} {244109}
  (\bibinfo {year} {2019})}\BibitemShut {NoStop}%
\bibitem [{\citenamefont {Mulvihill}\ \emph
  {et~al.}(2019{\natexlab{a}})\citenamefont {Mulvihill}, \citenamefont
  {Schubert}, \citenamefont {Sun}, \citenamefont {Dunietz},\ and\ \citenamefont
  {Geva}}]{Mulvihill2019}%
  \BibitemOpen
  \bibfield  {author} {\bibinfo {author} {\bibfnamefont {E.}~\bibnamefont
  {Mulvihill}}, \bibinfo {author} {\bibfnamefont {A.}~\bibnamefont {Schubert}},
  \bibinfo {author} {\bibfnamefont {X.}~\bibnamefont {Sun}}, \bibinfo {author}
  {\bibfnamefont {B.~D.}\ \bibnamefont {Dunietz}}, \ and\ \bibinfo {author}
  {\bibfnamefont {E.}~\bibnamefont {Geva}},\ }\href {\doibase
  10.1063/1.5055756} {\bibfield  {journal} {\bibinfo  {journal} {J. Chem.
  Phys.}\ }\textbf {\bibinfo {volume} {150}},\ \bibinfo {pages} {034101}
  (\bibinfo {year} {2019}{\natexlab{a}})}\BibitemShut {NoStop}%
\bibitem [{\citenamefont {Mulvihill}\ \emph
  {et~al.}(2019{\natexlab{b}})\citenamefont {Mulvihill}, \citenamefont {Gao},
  \citenamefont {Liu}, \citenamefont {Schubert}, \citenamefont {Dunietz},\ and\
  \citenamefont {Geva}}]{Mulvihill2019a}%
  \BibitemOpen
  \bibfield  {author} {\bibinfo {author} {\bibfnamefont {E.}~\bibnamefont
  {Mulvihill}}, \bibinfo {author} {\bibfnamefont {X.}~\bibnamefont {Gao}},
  \bibinfo {author} {\bibfnamefont {Y.}~\bibnamefont {Liu}}, \bibinfo {author}
  {\bibfnamefont {A.}~\bibnamefont {Schubert}}, \bibinfo {author}
  {\bibfnamefont {B.~D.}\ \bibnamefont {Dunietz}}, \ and\ \bibinfo {author}
  {\bibfnamefont {E.}~\bibnamefont {Geva}},\ }\href {\doibase
  10.1063/1.5110891} {\bibfield  {journal} {\bibinfo  {journal} {The Journal of
  Chemical Physics}\ }\textbf {\bibinfo {volume} {151}},\ \bibinfo {pages}
  {074103} (\bibinfo {year} {2019}{\natexlab{b}})}\BibitemShut {NoStop}%
\bibitem [{\citenamefont {Mulvihill}\ \emph {et~al.}(2021)\citenamefont
  {Mulvihill}, \citenamefont {Lenn}, \citenamefont {Gao}, \citenamefont
  {Schubert}, \citenamefont {Dunietz},\ and\ \citenamefont
  {Geva}}]{Mulvihill2021}%
  \BibitemOpen
  \bibfield  {author} {\bibinfo {author} {\bibfnamefont {E.}~\bibnamefont
  {Mulvihill}}, \bibinfo {author} {\bibfnamefont {K.~M.}\ \bibnamefont {Lenn}},
  \bibinfo {author} {\bibfnamefont {X.}~\bibnamefont {Gao}}, \bibinfo {author}
  {\bibfnamefont {A.}~\bibnamefont {Schubert}}, \bibinfo {author}
  {\bibfnamefont {B.~D.}\ \bibnamefont {Dunietz}}, \ and\ \bibinfo {author}
  {\bibfnamefont {E.}~\bibnamefont {Geva}},\ }\href {\doibase
  10.1063/5.0051101} {\bibfield  {journal} {\bibinfo  {journal} {Journal of
  Chemical Physics}\ }\textbf {\bibinfo {volume} {154}},\ \bibinfo {pages}
  {204109} (\bibinfo {year} {2021})}\BibitemShut {NoStop}%
\bibitem [{\citenamefont {Mulvihill}\ and\ \citenamefont
  {Geva}(2021)}]{Mulvihill2021b}%
  \BibitemOpen
  \bibfield  {author} {\bibinfo {author} {\bibfnamefont {E.}~\bibnamefont
  {Mulvihill}}\ and\ \bibinfo {author} {\bibfnamefont {E.}~\bibnamefont
  {Geva}},\ }\href {\doibase 10.1021/acs.jpcb.1c05719} {\bibfield  {journal}
  {\bibinfo  {journal} {Journal of Physical Chemistry B}\ }\textbf {\bibinfo
  {volume} {125}},\ \bibinfo {pages} {9834} (\bibinfo {year}
  {2021})}\BibitemShut {NoStop}%
\bibitem [{\citenamefont {Mulvihill}\ and\ \citenamefont
  {Geva}(2022)}]{Mulvihill2022}%
  \BibitemOpen
  \bibfield  {author} {\bibinfo {author} {\bibfnamefont {E.}~\bibnamefont
  {Mulvihill}}\ and\ \bibinfo {author} {\bibfnamefont {E.}~\bibnamefont
  {Geva}},\ }\href {\doibase 10.1063/5.0078040} {\bibfield  {journal} {\bibinfo
   {journal} {The Journal of Chemical Physics}\ }\textbf {\bibinfo {volume}
  {156}},\ \bibinfo {pages} {044119} (\bibinfo {year} {2022})}\BibitemShut
  {NoStop}%
\bibitem [{\citenamefont {Miller}\ and\ \citenamefont
  {White}(1986)}]{Miller1986}%
  \BibitemOpen
  \bibfield  {author} {\bibinfo {author} {\bibfnamefont {W.~H.}\ \bibnamefont
  {Miller}}\ and\ \bibinfo {author} {\bibfnamefont {K.~A.}\ \bibnamefont
  {White}},\ }\href {\doibase 10.1063/1.450655} {\bibfield  {journal} {\bibinfo
   {journal} {J. Chem. Phys.}\ }\textbf {\bibinfo {volume} {84}},\ \bibinfo
  {pages} {5059} (\bibinfo {year} {1986})}\BibitemShut {NoStop}%
\bibitem [{\citenamefont {Mahan}(1990)}]{MahanBook}%
  \BibitemOpen
  \bibfield  {author} {\bibinfo {author} {\bibfnamefont {G.~D.}\ \bibnamefont
  {Mahan}},\ }\href@noop {} {\emph {\bibinfo {title} {{Many-Particle
  Physics}}}}\ (\bibinfo  {publisher} {Plenum Press},\ \bibinfo {year}
  {1990})\BibitemShut {NoStop}%
\bibitem [{\citenamefont {Haug}\ and\ \citenamefont {Jauho}(2008)}]{HaugJauho}%
  \BibitemOpen
  \bibfield  {author} {\bibinfo {author} {\bibfnamefont {H.~J.~W.}\
  \bibnamefont {Haug}}\ and\ \bibinfo {author} {\bibfnamefont {A.-P.}\
  \bibnamefont {Jauho}},\ }\href@noop {} {\emph {\bibinfo {title} {{Quantum
  Kinetics in Transport and Optics of Semiconductors}}}},\ \bibinfo {edition}
  {2nd}\ ed.\ (\bibinfo  {publisher} {Springer},\ \bibinfo {address} {Berlin},\
  \bibinfo {year} {2008})\BibitemShut {NoStop}%
\bibitem [{\citenamefont {Newns}(1969)}]{Newns1969}%
  \BibitemOpen
  \bibfield  {author} {\bibinfo {author} {\bibfnamefont {D.~M.}\ \bibnamefont
  {Newns}},\ }\href {\doibase 10.1103/PhysRev.178.1123} {\bibfield  {journal}
  {\bibinfo  {journal} {Phys. Rev.}\ }\textbf {\bibinfo {volume} {178}},\
  \bibinfo {pages} {1123} (\bibinfo {year} {1969})},\ \Eprint
  {http://arxiv.org/abs/arXiv:1011.1669v3} {arXiv:arXiv:1011.1669v3}
  \BibitemShut {NoStop}%
\bibitem [{\citenamefont {Smith}\ and\ \citenamefont
  {Hynes}(1993)}]{smithElectronicFrictionElectron1993}%
  \BibitemOpen
  \bibfield  {author} {\bibinfo {author} {\bibfnamefont {B.~B.}\ \bibnamefont
  {Smith}}\ and\ \bibinfo {author} {\bibfnamefont {J.~T.}\ \bibnamefont
  {Hynes}},\ }\href {\doibase 10.1063/1.465843} {\bibfield  {journal} {\bibinfo
   {journal} {The Journal of Chemical Physics}\ }\textbf {\bibinfo {volume}
  {99}},\ \bibinfo {pages} {6517} (\bibinfo {year} {1993})}\BibitemShut
  {NoStop}%
\bibitem [{\citenamefont {Boroda}\ and\ \citenamefont
  {Voth}(1996)}]{Boroda1996a}%
  \BibitemOpen
  \bibfield  {author} {\bibinfo {author} {\bibfnamefont {Y.~G.}\ \bibnamefont
  {Boroda}}\ and\ \bibinfo {author} {\bibfnamefont {G.~A.}\ \bibnamefont
  {Voth}},\ }\href {\doibase 10.1063/1.471274} {\bibfield  {journal} {\bibinfo
  {journal} {J. Chem. Phys.}\ }\textbf {\bibinfo {volume} {104}},\ \bibinfo
  {pages} {6168} (\bibinfo {year} {1996})}\BibitemShut {NoStop}%
\bibitem [{\citenamefont {Thoss}, \citenamefont {Kondov},\ and\ \citenamefont
  {Wang}(2004)}]{Thoss2004a}%
  \BibitemOpen
  \bibfield  {author} {\bibinfo {author} {\bibfnamefont {M.}~\bibnamefont
  {Thoss}}, \bibinfo {author} {\bibfnamefont {I.}~\bibnamefont {Kondov}}, \
  and\ \bibinfo {author} {\bibfnamefont {H.}~\bibnamefont {Wang}},\ }\href
  {\doibase 10.1016/j.chemphys.2004.06.008} {\bibfield  {journal} {\bibinfo
  {journal} {Chem. Phys.}\ }\textbf {\bibinfo {volume} {304}},\ \bibinfo
  {pages} {169} (\bibinfo {year} {2004})}\BibitemShut {NoStop}%
\bibitem [{\citenamefont {Hillery}\ \emph {et~al.}(1984)\citenamefont
  {Hillery}, \citenamefont {O'Connell}, \citenamefont {Scully},\ and\
  \citenamefont {Wigner}}]{Hillery1984a}%
  \BibitemOpen
  \bibfield  {author} {\bibinfo {author} {\bibfnamefont {M.}~\bibnamefont
  {Hillery}}, \bibinfo {author} {\bibfnamefont {R.~F.}\ \bibnamefont
  {O'Connell}}, \bibinfo {author} {\bibfnamefont {M.~O.}\ \bibnamefont
  {Scully}}, \ and\ \bibinfo {author} {\bibfnamefont {E.~P.}\ \bibnamefont
  {Wigner}},\ }\href {\doibase 10.1016/0370-1573(84)90160-1} {\bibfield
  {journal} {\bibinfo  {journal} {Phys. Rep.}\ }\textbf {\bibinfo {volume}
  {106}},\ \bibinfo {pages} {121} (\bibinfo {year} {1984})}\BibitemShut
  {NoStop}%
\bibitem [{\citenamefont {Imre}\ \emph {et~al.}(1967)\citenamefont {Imre},
  \citenamefont {{\"{O}}zizmir}, \citenamefont {Rosenbaum},\ and\ \citenamefont
  {Zweifel}}]{Imre1967}%
  \BibitemOpen
  \bibfield  {author} {\bibinfo {author} {\bibfnamefont {K.}~\bibnamefont
  {Imre}}, \bibinfo {author} {\bibfnamefont {E.}~\bibnamefont {{\"{O}}zizmir}},
  \bibinfo {author} {\bibfnamefont {M.}~\bibnamefont {Rosenbaum}}, \ and\
  \bibinfo {author} {\bibfnamefont {P.~F.}\ \bibnamefont {Zweifel}},\ }\href
  {\doibase 10.1063/1.1705323} {\bibfield  {journal} {\bibinfo  {journal} {J.
  Math. Phys.}\ }\textbf {\bibinfo {volume} {8}},\ \bibinfo {pages} {1097}
  (\bibinfo {year} {1967})}\BibitemShut {NoStop}%
\end{thebibliography}
\end{document}